\documentclass[acmtog,nonacm,theauthorversion]{acmart}
\usepackage{booktabs} % For formal tables
\usepackage{xspace}

\usepackage{subcaption}

\usepackage{marvosym}
\usepackage{hyperref}       % hyperlinks
\usepackage{url}            % simple URL typesetting
\usepackage{booktabs}       % professional-quality tables
\usepackage{amsfonts}       % blackboard math symbols
\usepackage{nicefrac}       % compact symbols for 1/2, etc.
\usepackage{microtype}      % microtypography
\usepackage{xcolor}         % colors
\usepackage{amsmath}
\usepackage{blindtext}
\usepackage{graphicx}
\usepackage{epsfig}
\usepackage[ruled,vlined,linesnumbered]{algorithm2e}
\usepackage{tabularx}
\usepackage{enumitem}
\usepackage{color}

\usepackage{makecell}
\usepackage{multirow}

\usepackage{float}

\usepackage[normalem]{ulem}
\useunder{\uline}{\ul}{}

\usepackage{listings}

\usepackage[frozencache,cachedir=.]{minted}
\usepackage{siunitx}

\usepackage{mathtools}

\DeclarePairedDelimiter\floor{\lfloor}{\rfloor}

\SetAlFnt{\small}
\SetAlCapFnt{\small}
\SetAlCapNameFnt{\small}
\SetAlCapHSkip{0pt}
\SetKwInput{KwInput}{Input}                % Set the Input
\SetKwInput{KwOutput}{Output}

\makeatletter
\DeclareRobustCommand\onedot{\futurelet\@let@token\@onedot}
\def\@onedot{\ifx\@let@token.\else.\null\fi\xspace}

\def\eg{\emph{e.g}\onedot} 
\def\ie{\emph{i.e}\onedot}

\makeatother

\setcopyright{acmlicensed}
\acmJournal{TOG}
\acmYear{2025} \acmVolume{44} \acmNumber{6} \acmArticle{243} \acmMonth{12}\acmDOI{10.1145/3763358}

\begin{document}
% Title portion
\title[Large-Area Fabrication-Aware Computational Diffractive Optics]{Large-Area Fabrication-Aware Computational Diffractive Optics}

% DO NOT ENTER AUTHOR INFORMATION FOR ANONYMOUS TECHNICAL PAPER SUBMISSIONS TO SIGGRAPH 2019!
\author{Kaixuan Wei}
\authornote{indicates joint first authorship.}
%\orcid{1234-5678-9012-3456}
\affiliation{%
\institution{King Abdullah University of Science and Technology}
 \country{Saudi Arabia}
 }
\email{kaixuan.wei@kaust.edu.sa}

\author{Hector A. Jimenez-Romero}
\authornotemark[1]
%\orcid{1234-5678-9012-3456}
\affiliation{%
\institution{King Abdullah University of Science and Technology}
 \country{Saudi Arabia}
 }
\email{hector.jimenezromero@kaust.edu.sa}

\author{Hadi Amata}
\authornotemark[1]
%\orcid{1234-5678-9012-3456}
\affiliation{%
\institution{King Abdullah University of Science and Technology}
 \country{Saudi Arabia}
 }
\email{hadi.amata@kaust.edu.sa}

\author{Jipeng Sun}
%\orcid{1234-5678-9012-3456}
\affiliation{%
\institution{Princeton University}
\country{United States of America}
 }
\email{jipeng.sun@princeton.edu}

\author{Qiang Fu}
%\orcid{1234-5678-9012-3456}
\affiliation{%
\institution{King Abdullah University of Science and Technology}
 \country{Saudi Arabia}
 }
\email{qiang.fu@kaust.edu.sa}

\author{Felix Heide}
%\orcid{1234-5678-9012-3456}
\affiliation{%
 \institution{Princeton University}
 \country{United States of America}
 }
\email{fheide@cs.princeton.edu}

\author{Wolfgang Heidrich}
%\orcid{1234-5678-9012-3456}
\affiliation{%
 \institution{King Abdullah University of Science and Technology}
 \country{Saudi Arabia}
 }
\email{wolfgang.heidrich@kaust.edu.sa}

\authorsaddresses{%
Authors' addresses: \Letter Kaixuan Wei, kaixuan.wei@kaust.edu.sa, KAUST, Saudi Arabia; %
Hector A. Jimenez-Romero, hector.jimenezromero@kaust.edu.sa, KAUST, Saudi Arabia; %
Hadi Amata, hadi.amata@kaust.edu.sa, KAUST, Saudi Arabia; %
Jipeng Sun, jipeng.sun@princeton.edu, Princeton University, USA; %
Qiang Fu, qiang.fu@kaust.edu.sa, KAUST, Saudi Arabia;
Felix Heide, fheide@princeton.edu, Princeton University, USA; %
Wolfgang Heidrich, wolfgang.heidrich@kaust.edu.sa, KAUST, Saudi Arabia. %
}

\renewcommand\shortauthors{Wei, K. et al}

% formatting stuff

\definecolor{Gray}{rgb}{0.5,0.5,0.5}
\definecolor{darkblue}{rgb}{0,0,0.7}
\definecolor{orange}{rgb}{1,.5,0} % something readable but different from todo
\definecolor{red}{rgb}{1,0,0} % something readable but different from todo

\newcommand{\revise}[1]{{{#1}}}

\newcommand{\MATTHIAS}[1]{\textcolor{red}{[Matthias: #1]}}
\newcommand{\JRK}[1]{\revise{JRK: #1}}
\newcommand{\FELIX}[1]{\revise{Felix: #1}}
\newcommand{\IGNORE}[1]{}

\newcommand{\PROXIMAL}{ProxImaL~}

\newcommand{\secref}[1]{Sec.~\ref{#1}}
\newcommand{\figref}[1]{Fig.~\ref{#1}}
\newcommand{\tabref}[1]{Table~\ref{tab:#1}}

\newcommand{\heading}[1]{\noindent\textbf{#1}}
\newcommand{\note}[1]{{\em{\textcolor{orange}{#1}}}}
\newcommand{\todo}[1]{{\textcolor{red}{TODO: #1}}}
\newcommand{\kaixuan}[1]{{\textcolor{blue}{Kaixuan: #1}}}
\newcommand{\WH}[1]{{\textcolor{teal}{WH: #1}}}
\newcommand{\lzq}[1]{{\textcolor{blue}{Zeqiang: #1}}}
\newcommand{\changed}[1] {#1} %{{\textcolor{red}{#1}}}
\newcommand{\place}[1]{ \begin{itemize}\item\textcolor{darkblue}{#1}\end{itemize}}
\newcommand{\de}{\mathrm{d}}

\newcommand{\specialcell}[2][c]{%
  \begin{tabular}[#1]{@{}c@{}}#2\end{tabular}}

% \newcommand{\code}[1]{{\small \textbf{\texttt{#1}}}}

% math stuff
\newcommand{\Mat}[1]    {{\ensuremath{\mathbf{\uppercase{#1}}}}} %Matrix
\newcommand{\Vect}[1]   {{\ensuremath{\mathbf{\lowercase{#1}}}}} %Vector
\newcommand{\Vari}[1]   {{\ensuremath{\mathbf{\lowercase{#1}}}}} %Vector
\newcommand{\Id}				{\mathbb{I}} %Identity matrix
\newcommand{\Diag}[1] 	{\operatorname{diag}\left({ #1 }\right)} %Diagonalized matrix
\newcommand{\Opt}[1] 	  {{#1}_{\text{opt}}} %Optimal point of an optimization
\newcommand{\CC}[1]			{{#1}^{*}} %Convex conjugate
\newcommand{\Op}[1]     {\Mat{#1}} %Operator
\newcommand{\mini}[1] {{\operatorname{argmin}}_{#1} \: \: } %Minimize w.r.t.
\newcommand{\minimize}[1] {\underset{{#1}}{\operatorname{argmin}} \: \: } %Minimize w.r.t.
\newcommand{\maximize}[1] {\underset{{#1}}{\operatorname{argmax}} \: \: } %Maximize w.r.t.
\newcommand{\grad}      {\nabla}
\newcommand{\kron}{\otimes} %Pointwise multiplication

\newcommand{\gradt}     {\grad_\z}
\newcommand{\gradx}     {\grad_\x}
\newcommand{\step}      {\text{\textbf{step}}}
\newcommand{\prox}[1]   {\mathbf{prox}_{#1}}
\newcommand{\ind}[1]    {\operatorname{ind}_{#1}}
\newcommand{\proj}[1]   {\Pi_{#1}}
\newcommand{\pointmult}{\odot} %Pointwise multiplication

% notation for method
\newcommand{\conv}{\ast} %Convolution
\newcommand{\meas}{\Vect{y}}            % measurement vector
\newcommand{\Img}{I}                    % transient image
\newcommand{\img}{\Vect{i}}             % vectorized image
\newcommand{\x}{\Vect{x}}             % vectorized image
\newcommand{\z}{\Vect{z}}             % vectorized image
\newcommand{\p}{\Vect{p}}
\newcommand{\Splitvar}{T}                % dual var
\newcommand{\splitvar}{\Vect{t}}         % dual vectorized var
\newcommand{\Splitbasis}{J}                % dual basis
\newcommand{\splitbasis}{\Vect{j}}         % dual vectorized basis
\newcommand{\lagrangemult}{\Vect{\lambda}}         % dual vectorized volume
\newcommand{\var}{\Vect{x}}
\newcommand{\Kvar}{\Vect{z}}
\newcommand{\pvar}{\Vect{v}}
\newcommand{\eps}{\epsilon}

%Fourier notation
\newcommand{\FT}[1]			{\boldsymbol{\mathcal{F}}\left( {#1} \right)} %Fourier transform
\newcommand{\IFT}[1]
{\boldsymbol{\mathcal{F}}^{-1}\left( {#1} \right)} %Inverse Fourier transform

%Optimization
\newcommand{\func}{f}
\newcommand{\fMat}{\Mat{K}}

\newcommand{\avar}{\Vari{v}}
\newcommand{\aspvar}{\Vari{z}}

\newcommand{\maskM}{\Mat{M}}    % mask
\newcommand{\blurM}{\Mat{D}}    % blur kernel / convolution matrix
\newcommand{\diagM}{\Delta}     % generic diagonal matrix
\newcommand{\gradM}{\nabla}     % gradient matrix
\newcommand{\idM}{\Mat{I}}      % identity matrix
\newcommand{\orthM}{\Mat{Q}}    % generic orthogonal matrix
\newcommand{\genericMA}{\Mat{A}} % generic matrix
\newcommand{\genericMB}{\Mat{B}} % a second generic matrix
\newcommand{\fourierM}{\mathcal{F}} % Fourier matrix
\newcommand{\vp}{\Vari{v}}

\newcommand{\Pen}      		{F} %Penalty function
\newcommand{\cardset}     {\mathcal{C}}
\newcommand{\Dat}      		{G} %Dataterm function
\newcommand{\Reg}      		{\Gamma} %Regularizer spatial

% New stuff.
\newcommand{\argmin}{\mathop{\rm argmin}}
\newcommand{\reals}{\mathbb{R}}%{{\mbox{\bf R}}}
\newcommand{\integers}{\mathbb{Z}}%{{\mbox{\bf Z}}}
\newcommand{\epi}{\mathop{\bf epi}}
\newcommand{\complex}{\mathbb{C}}%{{\mbox{\bf C}}}
\newcommand{\symm}{{\mbox{\bf S}}}  % symmetric matrices

% text abbrevs
% \newcommand{\eg}{{\it e.g.}}
% \newcommand{\ie}{{\it i.e.}}
% \newcommand{\etc}{{\it etc.}}

\newcommand{\aka}{{\it a.k.a.~}}

\def\shortname{$\nabla$-Prox}

\newcommand{\subjectTo}{s.t.} 

\newcommand{\wei}[1]{\textcolor{red}{{[Wei: #1]}}}

\begin{teaserfigure}
  \centering
  \includegraphics[width=.99\textwidth]{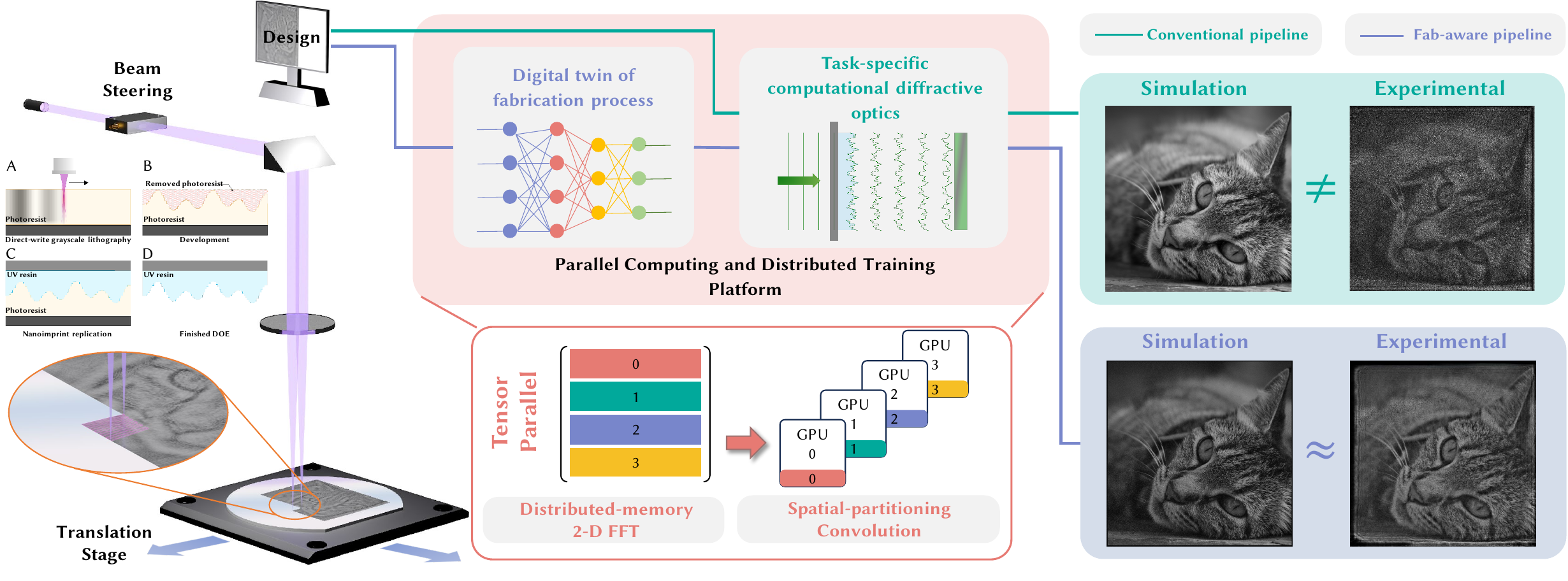}
  \caption{\textbf{Fabrication-aware, End-to-end Optimization for Large-area Diffractive Optics.} We propose a fabrication-aware design method for diffractive optical elements fabricated by (left) direct-write grayscale lithography with nanoimprint replication (see the inset figures A-D for a step-by-step illustration) suited for inexpensive mass production. Enabled by tensor-parallel computing routines, our method jointly considers the fabrication 3-D geometry deformation and the downstream task-specific computational diffractive optics design. This combination of techniques allows for experimental findings with favorable quality to all tested existing methods, specifically closing the design-to-manufacturing gap in existing approaches.
  }
  \label{fig:teaser}
\end{teaserfigure}
 
\begin{abstract}

Differentiable optics, as an emerging paradigm that jointly optimizes
optics and (optional) image processing algorithms, has made many
innovative optical designs possible across a broad range of imaging and display
applications. Many of these systems utilize diffractive optical
components for holography, PSF engineering, or wavefront
shaping. Existing approaches have, however, mostly remained limited to laboratory prototypes, owing to a large quality gap between simulation and manufactured devices.

We aim at lifting the fundamental technical barriers to the practical use of learned diffractive optical systems. To this end, we propose a fabrication-aware design pipeline for diffractive optics fabricated by direct-write grayscale lithography followed by replication with nano-imprinting, which is directly suited for inexpensive mass-production of large area designs. We propose a super-resolved
neural lithography model that can accurately predict the 3D geometry
generated by the fabrication process. This model can be seamlessly
integrated into existing differentiable optics frameworks, enabling
fabrication-aware, end-to-end optimization of computational optical
systems. To tackle the computational challenges, we also devise tensor-parallel compute framework centered on distributing large-scale FFT computation across many GPUs.

As such, we demonstrate large scale
diffractive optics designs up to 32.16\,mm $\times$ 21.44\,mm, simulated on grids of up to 128,640 by 85,760 feature points. We find adequate agreement between simulation and fabricated prototypes for applications such as holography and PSF engineering. We also achieve high image quality from an imaging system comprised only of a single diffractive optical element, with images processed only by a one-step inverse filter utilizing the simulation PSF. We believe our findings lift the fabrication limitations for real-world applications of diffractive optics and differentiable optical design.

%Moreover, to overcome the intrinsic memory bottlenecks in simulating-and-optimizing generic large-aperture optical systems, we realize a differentiable distributed-memory fast Fourier transform (D$^2$FFT) which supports the large-scale wave propagation in tensor/model parallelism beyond the naive data parallelism. Paired with a tensor-parallel convolutional neural network as the neural lithography model, this allows for the high-fidelity optimization of diffractive optical elements with unprecedented scale/resolution, discretized by a 100,000 $\times$ 100,000 (\ie, ten billions) pixel array, efficiently distributed across computing nodes for example. Extensive experiments over three real-world applications, including computational display holography, beam shaping, and broadband color imaging, collectively demonstrate the effectiveness of the proposed approach in designing and fabricating high-performance diffractive optical systems.  

\end{abstract}

\begin{CCSXML}
<ccs2012>
   <concept>
       <concept_id>10010147.10010341.10010342.10010343</concept_id>
       <concept_desc>Computing methodologies~Modeling methodologies</concept_desc>
       <concept_significance>500</concept_significance>
       </concept>
<concept>
    <concept_id>10010405.10010432.10010439</concept_id>
    <concept_desc>Applied computing~Engineering</concept_desc>
    <concept_significance>500</concept_significance>
</concept>       
 </ccs2012>
\end{CCSXML}

\ccsdesc[500]{Computing methodologies~Modeling methodologies}
\ccsdesc[500]{Applied computing~Engineering}
% \ccsdesc[500]{Computing methodologies~Computational photography; Regularization}
% \ccsdesc[300]{Mathematics of computing~Continuous optimization; Solvers}


\keywords{Computational optics, computational imaging, computational fabrication}

\maketitle

\section{Introduction}

Over the last decades, rapid advances in computational power, and
photodetection devices have enabled the emergence of computational
imaging and optics as a new way for designing optical
systems~\cite{mait_computational_2018,bhandari_computational_2022}. By
co-designing optics and image processing algorithms, these
computational systems can produce new forms of visual information, which
are otherwise difficult to capture by traditional optical
systems~\cite{nayar_computational_2006}. Notably,
diffractive optical elements (DOEs) are particularly suited for
computational optics as they can encode complex optical functions,
such as an arbitrary phase modulation, that are hard to achieve with
refractive optics. As a result, remarkable capabilities such as
snapshot high-dynamic-range \cite{sun_learning_2020}, extended
depth-of-field \cite{nehme_deepstorm3d_2020}, hyperspectral
\cite{shi2024learned} and monocular depth imaging
\cite{ikoma2021depth}, have been demonstrated in DOE-based
computational systems, where the complex DOE design problems are
addressed by the differentiable modeling of the wave optics system
\cite{sitzmann_end--end_2018}, and back-propagation-based optimization
popularized by the advent of deep learning (DL) era. The resulting
paradigm, coined deep optics or differentiable optics, has shown to be
a versatile tool to optimize computational diffractive, refractive
\cite{sun_end--end_2021,yang_curriculum_2024} or hybrid
refractive-diffractive \cite{yang_end--end_2024} optical systems, not
just for imaging but for near-eye display
\cite{chakravarthula_wirtinger_2019,peng_neural_2020}, with
extraordinary task-specific performance.  

Despite the seemingly striking results, these advanced learned
diffractive optical systems have largely remained limited to
laboratory prototypes, owing to the large quality discrepancies
between simulation and fabricated
devices~\cite{zhao2023close,shi2024split}. Unlike refractive optical
elements with smooth surfaces, DOEs rely on precise control of
micron-sized structures that can quickly vary across the design space,
thereby requiring sophisticated micro/nano-fabrication technology.

To understand the reason for the discrepancy between simulated designs
and fabricated prototypes, we need to consider a specific fabrication
process in some detail. Here we explain grayscale lithography with
direct laser writing (see Fig.~\ref{fig:teaser}, left)~\cite{grushina_direct-write_2019}; other
lithographic method such a two-photon
polymerization~\cite{wang_twophoton_2023} differ in the details but
require similar high-level considerations. In direct laser writing, a
laser beam scans across a wafer covered in thin layer photoresist in a
2D grid. Modulation of the beam intensity is used to create a
spatially varying exposure map (A). The first notable effect in this
process is an optical blur of the specified design, which can be
modeled as the convolution of the design pattern with the point spread
function of the laser beam and can, for example, result in corner
rounding for small rectangular features. Next, the local exposure level
induces a local chemical change in the photoresist, which can be
thought of as a nonlinear local transfer function. The next step (B) is
the development of the exposed resin, in which the developer
decomposes the photoresist, where the rate of decomposition depends on
the local exposure received by the resist. This is a complicated 3D
chemical interaction that finally reveals a height field
structure. Nanoimprinting can be used to transfer this shape into a UV
resin (C) to form the final DOE (D). Further shape distortions are
possible in this step, for example, due to volume shrinkage of the
resin during curing. For mass production, only the final imprinting
step needs to be repeated for each copy.

Inherent to the fabrication process is that the
fabricated DOE will differ from the target shape at scales \emph{smaller
than the intended feature size}. To accurately predict the optical
performance of a DOE, we not only need to have a precise digital twin
of the fabrication process, but also need to conduct the optical
simulation at substantially super-resolved resolutions compared to the
target feature size.

A related source of error is that most inverse design pipelines for
diffractive optics work on drastically \emph{undersampled grids} out of
computational necessity. Most early DOE design works use simulation
grids where one grid point equals one DOE feature. More recently,
simulations have started using a $2\times$ finer grid
(\eg,~\cite{yang_end--end_2024}). However, according to the Nyquist
limit any sample grid can only represent sinusoidal components up to
twice the grid pitch, whereas many holography and DOE designs rely on
sharp step edges between neighboring features. We experimentally
demonstrate this problem in Section~\ref{sec:neural-lithography} and
Figure~\ref{fig:motivation-example} by performing simulations on the
original feature grid resolution and on an $8\times$ super-resolved
grid with nearest neighbor upsampling to force box-shaped features,
highlighting drastic simulation differences even without accounting
for fabrication limitations.

Several strategies for dealing with the large simulation gap have been
introduced and are in common use: commercial product design often
relies on a tedious iterative process between design, fabrication, and
measurement~\cite{jang_design_2020}. In the holographic display
community, camera-in-the-loop systems have recently become
popular\revise{~\cite{peng_neural_2020,kavakli2021learned,choi2022time}}, whereas in end-to-end designed
computational imaging systems, it is common practice to fine-tune the
computational module with measurements from the as-fabricated
prototype
(\eg,~\cite{peng_learned_2019,shi2024split,chakravarthula_thin_2023}
etc.). However, closing the loop in this fashion is expensive and may
not be practical beyond lab prototypes.

In this work, we instead desire a fabrication-aware design process that
can accurately predict the final system in
open-loop simulation. The core of this approach is a digital twin for
the fabrication process in the form of a neural lithography model for
direct-write grayscale lithography followed by nanoimprinting. This
model can easily be incorporated into any existing diffractive design
pipeline. Unlike the recent neural lithography work by Zheng et
al.~\shortcite{zhao2023close}, we directly target a fabrication
process that is suitable for both large-scale designs as well as mass
fabrication. To facilitate the resulting large-scale design processes
and tackle the challenges posed by large memory requirements, we also
devise a parallelization toolbox for distributing large FFTs across
many GPUs. To demonstrate the efficacy and generalization capabilities of our approach, we validate the method for several applications:

\begin{itemize}
\item Computer-generated holograms produced open-loop (i.e. without camera in the
  loop) with substantially reduced noise and speckle, and excellent
  agreement between simulation and prototype. This includes a design
  of up to 32.16 mm $\times$ 21.44 mm in size, simulated on a grid of
  128,640 $\times$ 85,760 feature points on 16 A-100 GPUs.
\item We conduct a beam shaping experiment on the example of splitting an
  incident beam into a regular grid of output beams with controlled
  intensities, which can be used for downstream tasks such as laser
  material processing~\cite{kuang2013ultrafast} and 3-D
  sensing~\cite{yuan2021fast} or visual
  vibrometry~\cite{zhang2023impacts}.
\item We design a computational camera comprised of a single DOE and off-the-shelf sensor for broadband color imaging, again showing
  outstanding agreement between simulation and prototype, to the point
  that high image quality can be achieved by a one-step inverse
  filter utilizing the simulated PSF of the system. 
\end{itemize}

The last application confirms the ability to not only eliminate
fine-tuning of the image restoration module based on measured
characterization of the prototype, but also demonstrates that, with
 a design process that includes an accurate model of the manufacturing process, image restoration does not necessarily
require heavy deep neural networks but can utilize more lightweight
architectures that are compatible with the computational resources of
edge devices. We believe that this, by itself, is a major step forward in
improving the practicality of diffractive end-to-end design beyond lab
prototypes. 
Our code is publicly available at https://github.com/Vandermode/LAFA

\section{Related Work}

In the following, we review work on diffractive optical elements, computational lithography, and distributed training frameworks related to the method proposed in this paper.

% \paragraph{Diffractive Optical Elements.~} 
\subsection{Diffractive Optical Elements}
% Manipulating light via diffraction instead of refraction or reflection, the conceptualization of DOEs can be attributed to Fresnel who proposed a zone plate used to focus light in 19th century. It is until 1948, 
In 1948, Dennis Gabor introduced the optical holography, the first physical realization of DOEs through interference \cite{Gabor1948}. 
Since then, various forms of DOEs have been proposed, such as computer-generated holograms \cite{brown1966complex}, binary gratings \cite{collischon1994binary} and kinoform lens \cite{lesem1969kinoform}, to name a few---see \cite{zhang_diffractive_2023} for a comprehensive review of long history of DOEs. However, due to intrinsic difficulty in designing and manufacturing DOEs, their uses remain largely limited to well-controlled laboratory settings. Over the last decade, advances in computational imaging and the increasing capabilities of nano-fabrication technology have allowed researchers to revisit these conventional DOEs, as powerful optical encoders in hardware-software co-designed computational systems \revise{\cite{peng_computational_2015,peng_diffractive_2016,heide_encoded_2016,asif_flatcam_2017,boominathan_lensless_2016,sitzmann_end--end_2018,wu2019phasecam3d}}.  These methods have been successful in a wide array of applications, including full-spectral color imaging \cite{peng_diffractive_2016,heide_encoded_2016}, extended depth-of-field imaging \cite{tan_codedstereo_2021}, compressive lensless imaging \cite{asif_flatcam_2017,boominathan_phlatcam_2020}, single-shot hyperspectral \cite{shi2024learned} and depth imaging \cite{baek_single-shot_2021,shi2024split}, high-dynamic-range imaging \cite{metzler_deep_2020,sun_learning_2020}, seeing through obstructions \cite{shi_seeing_2022}, ultra-wide-angle holographic display \cite{tseng2024neural}, and computer vision tasks \cite{wei2024spatially}. Although successful, two fundamental technical obstacles remain: 1) the gap from design to manufacturing and 2) the challenge to design large-area devices at high fidelity, which are essential to compete with the widely used refractive optics. %In this work, we face these long-standing challenges directly and demonstrate fabrication-aware computational diffractive optics at unprecedented scale. 

% \paragraph{Micro/nano-Fabrication and Computational Lithography.~}
\subsection{Nano-Fabrication and Computational Lithography}
Optical lithography \cite{dill_optical_1975}, one of the key driving forces behind Moore's Law, has made tremendous progress over the last half a century \cite{moore1998cramming,mack_fifty_2011}. 
In parallel to resolution improvement via shorter-wavelength illuminator and higher numerical aperture (NA) optics \cite{bruning_optical_2007}, a body of work explores algorithmic resolution improvement---optical proximity correction (OPC) \cite{fung1997optical}, inverse lithography \cite{cecil_advances_2022,pang_inverse_2021}, or computational lithography \cite{lam_computation_2009,ma2011computational}. % in response to the increasing considerable technical and financial difficulties in physical system scaling \cite{schellenberg_resolution_2004,lam_computation_2009,ma2011computational,pang_inverse_2021,cecil_advances_2022}.
% Early works in computational lithography 
% Adoptions of this resolution enhancement technology evolve from early rule-based/model-based OPC \cite{fung1997optical,cobb1997experimental} that heuristically precompensates the photomask design based upon empirical or model-based knowledge of the photolithography process, to inverse lithography \cite{cecil_advances_2022} in recent years that solves a rigorous mathematical inverse problem with manufacture-induced constraints \cite{lam2010regularization}. 
% Albeit the successful demonstrations of these computational lithography techniques, 
% significant roadblocks to the broad production applications remain, such as the full-chip scalability and the manufacturability of the complex inverse-designed photomasks \cite{pang_inverse_2021}. 
% Moreover, 
These existing works are designed for the mask-based photolithography process tailored for 2-D binary patterns representative of integrated circuits, which unfortunately are not directly applicable to the fabrication of 3-D (2.5D) DOEs with continuously varying height profiles. 

Since the advent of commercially available 3-D micro/nano patterning techniques, such as the two-photon polymerization (TPP) lithography \cite{wang_twophoton_2023,wang_two-photon_2024} and the direct-write grayscale lithography \cite{grushina_direct-write_2019}, a number of works focused on the physical process modeling \cite{guney_estimation_2016,onanuga2019process,saha_effect_2017} and thereby the structure prediction and precompensation \cite{wang_toward_2020,lang_towards_2022,chevalier_rigorous_2021} for 3-D micro/nano fabrication. 
% These physical simulators show good aggreement between prediction and experimental measurements, and thus can be used to alleviate the structural deviations from the ideal design \cite{chevalier_rigorous_2021,lang_towards_2022}. 
These physical simulator-based approaches, however, largely rely on heuristics and iterative trial-and-error to precompensate the design. Recently, Zheng et al. \citeyearpar{zhao2023close} proposed neural lithography, a differentiable neural network (NN)-based fabrication simulator that enables joint optical design and fabrication correction end-to-end, automatically guaranteeing manufacturability. However, their approach is limited to low-throughput TPP lithography for micron-sized DOE patterns. Most recently, a concurrent work to ours \cite{xu_fabrication-integrated_2025}
proposed a differentiable model-based physical simulator for direct-write grayscale lithography. %that is integrated with DOE design in the end-to-end optimization. 
In contrast to our work with large-area (centimeter-scale) devices as
goal, they do not address the replication challenge
\cite{barcelo_nanoimprint_2016} while are limited in moderate-sized
(millimeter-scale) devices. We also demonstrate that our data-driven
neural model is more accurate than their simulation-based framework.

\begin{figure*}[!t]
    \centering
    \includegraphics[width=0.99\linewidth]{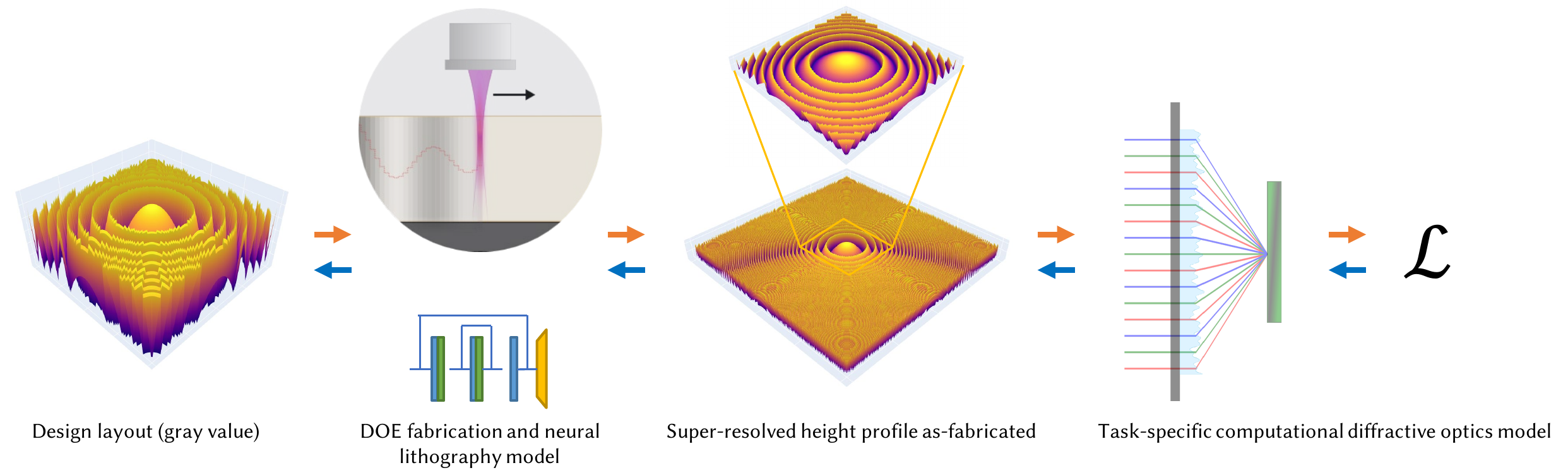}
    \caption{\textbf{Fabrication-aware Image Formation Model.} We illustrate the proposed image formation model (see text) for computational diffractive optics. With this differentiable model in hand, we optimize design layouts (\ie, inputs of the lithography machine) end-to-end informed by the proposed super-resolved neural lithography model, via backpropagation.  
    }
    \label{fig:image-formation}
\end{figure*}

% \paragraph{Parallel Computing and Distributed Training in Deep Learning.~}
\subsection{Parallel Computing and Distributed Training in Deep Learning}

% 3d Parallelism: data parallelism, tensor parallelism, pipeline parallelism
Our work also takes inspiration from recent advances in parallel computing and distributed training infrastructure~\cite{radford2019language,brown2020language,narayanan_efficient_2021}.
The most common parallelization strategy is data parallelism \cite{hillis1986data}, which distributes the data across different computing nodes and operate on the data in parallel. 
% For DL, this is amount to splitting the training data into mini-batches and distributing them across multiple GPUs \cite{krizhevsky2012imagenet}.  
However, this technique has a fundamental limitation in the model size it can tackle---the model must fit entirely on one worker. With the increasing size and complexity, NNs have approached the memory capacity of modern hardware accelerators. To overcome this bottleneck, \emph{model parallelism} \cite{shoeybi_megatron-lm_2020} has been proposed for training billion-parameter-scale LLMs, which includes  pipeline parallelism \cite{huang2019gpipe} and more general \emph{tensor parallelism} \cite{narayanan_efficient_2021}. Pipeline parallelism splits the NN pipeline into multiple stages across devices at the expense of the bubble overhead; Tensor parallelism is specialized for distributing specific atomic operators, such as the general matrix multiplication widely used in the Transformer block \cite{vaswani_attention_2017}. For computational optics, data parallelism supports scaling of wavelengths, incident angles, propagation distances, but cannot realize designs of large-area/scale DOEs discretized with billion-pixel-scale huge arrays \cite{sun2025collaborative}. As such, we leverage the tensor parallelism and devise the specialized computing routines for the large-area wave-optics model. \revise{This includes differentiable distributed-memory (D$^2$) FFT for free-space wave propagation as well as spatial-partitioning convolutional neural networks as the neural lithography model. 
Although the basic idea of distributed-memory FFT can be traced back to decades ago in the high-performance computing community \cite{gupta2002scalability} with implementations in CPU \cite{frigo2005design,pippig_pfft_2013}, GPU \cite{gholami_accfft_2016,ayala2022performance} or TPU \cite{lu_large-scale_2021} clusters based on Message Passing Interface standard (MPI) \cite{snir1998mpi}, none of these methods have been designed for differentiable wave optics simulation.}
Our implementation follows the generalized single program multiple data (GSPMD) programming model \cite{xu2021gspmd,shazeer2018mesh}, and thereby flexibly supports any tensor-dimensions to be split across any dimensions of a multi-dimensional mesh of processors---realizing arbitrary hybrid data and model (tensor) parallelism without painful code rewriting.

\section{Fabrication-aware End-to-end Design of Diffractive Optical Systems}

In this section, we first introduce our fabrication-aware image
formation model for computational diffractive optics (\figref{fig:image-formation}, Section
\ref{sec:image-formation}). Then, we describe the proposed
super-resolved neural lithography model as a differentiable digital
twin of the fabrication process (Section
\ref{sec:neural-lithography}). The proposed computational tools for
performing this large scale simulation and design tasks, \ie,
distributed-memory FFT and tensor-parallel convolution, for simulating
and optimizing large-area/scale wave optics are described in
Section~\ref{sec:parallelization}.

\subsection{Fabrication-aware Image Formation Model}
\label{sec:image-formation}

We model the DOE in our forward model as a phase profile
$\Phi_\lambda$ (with respect to a nominal wavelength $\lambda$) or,
equivalently, as a height map $h$, which the relationship
\begin{align}
    \Phi_\lambda(x, y) = \frac{2\pi}{\lambda} (n_\lambda-1) h(x, y),
\end{align}
where $n_\lambda$ is the wavelength-dependent refractive index of the DOE material (such as resins). Given an incident wave, such as a plane or spherical wave field, $E^\text{in}_\lambda(x, y)$, the modulated wavefront at the DOE plane is given by
\begin{align}
    E^\text{doe}_\lambda(x, y) = E^\text{in}_\lambda(x, y) e^{i \Phi_\lambda(x, y)}.
\end{align}
Then, the destination field $E^\text{dest}_\lambda$ in the sensor plane is given via the Rayleigh-Sommerfeld diffraction integral \cite{goodman2005introduction}, implemented by a numerical diffraction propagation method such as the angular spectrum method (ASM) \cite{matsushima_band-limited_2009,matsushima_shifted_2010,ritter_modified_2014,zhang_frequency_2020}, \ie, 
\begin{align}
    E^\text{dest}_\lambda = \mathcal{F}^{-1} \left\{ \mathcal{F} \left\{ E^\text{doe}_\lambda \right\} \otimes \mathrm{H}_\lambda  \right\},
    \label{eq:asm}
\end{align}
where $\mathcal{F} \{\cdot\}$ represents the fast Fourier transform (FFT), $\otimes$ denotes the Hadamard (elementwise) product, $\mathrm{H}_\lambda$ is the transfer function associated with the propagation model. 
Finally, the task-specific optimization of the diffractive optical system is posed as
\begin{align}
    h^* = \argmin_{h} \sum_{\lambda} \mathcal{L}_p \left( \| E^\text{dest}_\lambda \left(h \right) \|^2 \right), 
    \label{eq:general-opt-obj}
\end{align}
where $\mathcal{L}_p$ is a task-specific penalty/loss function defined
using the intensity $I_\lambda = \| E^\text{dest}_\lambda \|^2$ of the
destination field, \ie, point spread functions (PSF) for
imaging/sensing, or coherent hologram images for holography
\cite{chakravarthula_wirtinger_2019}. Note that the overall
optimization objective may have additional parameters such as
wavelengths, incident angles, object distances, and other
propagation-related parameters. These can result in a large number of
simulations according to Eq.\eqref{eq:asm}. Training data and
reconstruction methods can also be incorporated for joint optimization
of optics and image processing \cite{sitzmann_end--end_2018}. Here, we
only include the wavelength $\lambda$ dependency in
\eqref{eq:general-opt-obj} for simplicity without loss of generality.

The aforementioned image formation model and the derived optimization objective are general enough to cover a broad array of works in computational diffractive optics \cite{shi2024split}. However, most existing work largely ignores the manufacturing process and assume the designed DOE can be fabricated as it is, presuming a perfect indentity mapping between design layouts and manufactured devices. In reality, due to the sophisticated photolithography process, 3-D optical proximity effects as well as the complex photochemical interaction render significant deviations from design to manufacturing. We model these deviations with a lithography model $\mathcal{G} \{ \cdot \}$, akin to \cite{zhao2023close}, a surrogate of the fabrication process, which maps the design layout $l$ to the device parameterized by $h = \mathcal{G} \{ l \}$.  As such, the objective \eqref{eq:general-opt-obj} is reformulated as
\begin{align}
    l^* = \argmin_{l} \sum_{\lambda} \mathcal{L}_p \left( \| E^\text{dest}_\lambda \left( \mathcal{G} \left\{ l \right\} \right) \|^2 \right),
    \label{eq:fab-aware-opt}
\end{align}
which can be solved by back-propagation and gradient-based optimization \cite{kingma2014adam} if all components including the lithography model $\mathcal{G}$ are implemented as differentiable operators.

\subsection{Super-resolved Neural Lithography Model}
\label{sec:neural-lithography}

\begin{figure}[t]
    \centering
    \includegraphics[width=0.99\linewidth]{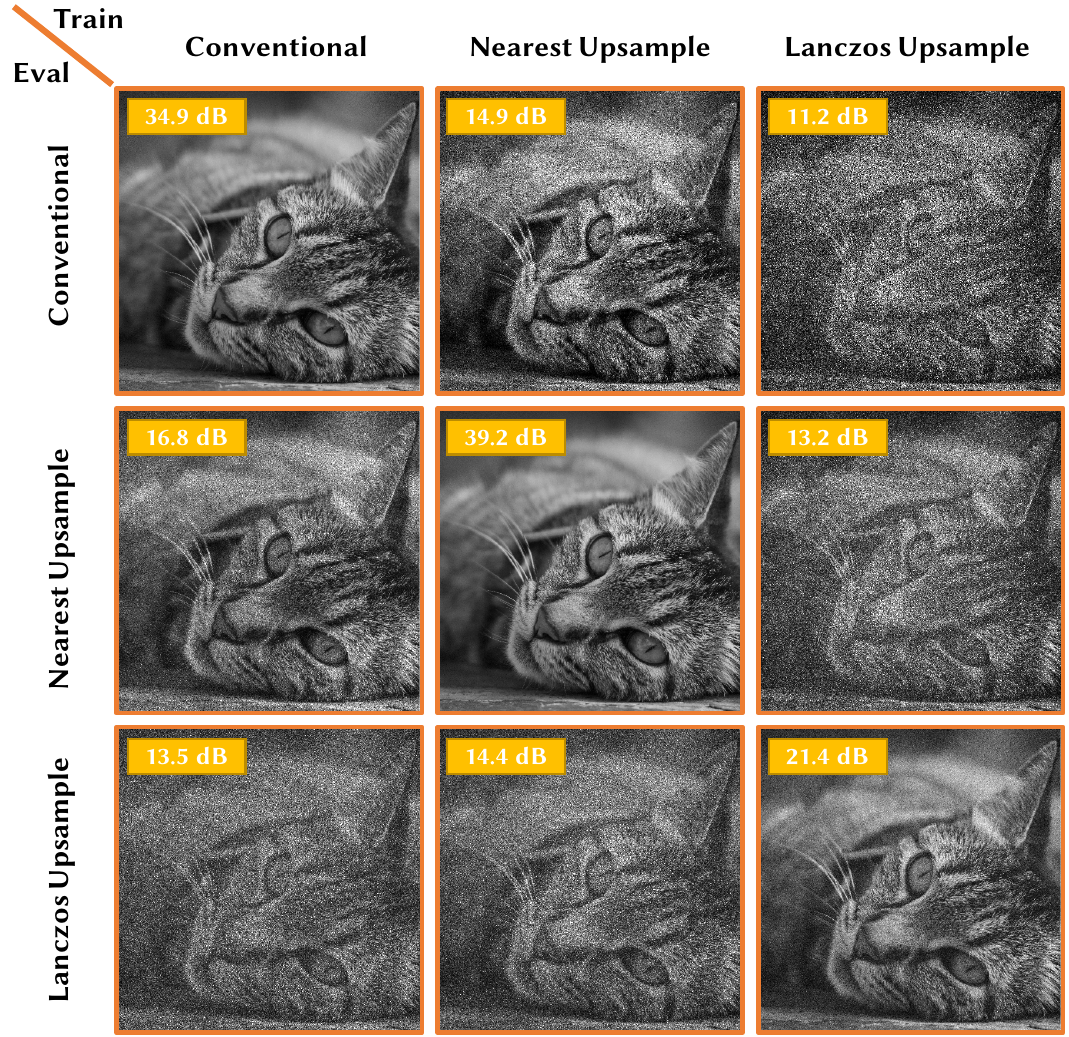}
    \caption{\textbf{Toy Example of the ``Fabrication Interpolation Kernel''}. We train and evaluate DOEs for computer-generated 2-D hologram under different settings, including 1) the conventional design at \SI{2}{\micro\metre}-spacing grid; 2) design with nearest upsampling in DOE plane at 250 nm-spacing grid (\ie, 8$\times$ upsampling) that meets the $\frac{\lambda}{2}$-spacing requirement of the Nyquist sampling theorem; 3) design with Lanczos upsampling in the DOE plane at 250 nm-spacing grid. Resulting PSNRs are shown in the top-left corner for each image. \revise{A comprehensive ablation study is provided in the Supplementary Material.}  }
    \label{fig:motivation-example}
\end{figure}

Next, we motivate core concepts of the method with a toy example that demonstrates the need for fabrication-aware optimization. Then we introduce the fabrication pipeline towards large-area DOE device manufacturing under the mass-production-ready setting. Finally, we detail the calibration methods to build the proposed neural lithography model, including the contrast curve calibration and neural lithography learning.

\paragraph{Toy Example.~} A \revise{3.6$\times$3.6 mm$^2$} DOE with a feature size of 2
$\mu$m is optimized to generate a 2-D hologram image under coherent
(520.6 nm wavelength) laser illumination. We use the Adam optimizer \cite{kingma2014adam} to solve \eqref{eq:general-opt-obj}, where the penalty $\mathcal{L}_p$ is defined as a combination of a scale-invariant mean square error loss $\mathcal{L}_{\text{si-mse}}$ and an energy regularization $\mathcal{L}_{\text{energy}}$ \footnote{Detailed definitions of the loss functions are given in the Supplementary Material} as
\begin{align}
    \mathcal{L}_p = \mathcal{L}_{\text{si-mse}} + \beta \mathcal{L}_{\text{energy}},
    \label{eq:hologram-loss}
\end{align}
where $\beta$ is empirically set as $5\times 10^{-3}$. The DOE height map is
randomly initialized, and the learning rate is initially set as
$10^{-2}$, which follows a cosine schedule that decays to zero in 3000
iterations.  Such an optimization leads to almost perfect 2-D hologram
reconstruction (34.9 dB in PSNR) evaluated under this conventional
setting (see the left-top image in
Fig.~\ref{fig:motivation-example}). However, as explained in the
introduction, the optical system simulation at \SI{2}{\micro\metre}-spacing grid violates the Nyquist sampling theorem \cite{smith_scientist_1999}, which inevitably results in aliasing artifacts degrading performance in reality. The zero-order interpolation, \ie, nearest upsampling is typically employed \cite{kuo_multisource_2023,tseng2024neural} to alleviate this issue, but this assumes perfect flat feature structure can be reliably manufactured. Other smooth/low-pass interpolation kernels such as the Lanczos-3 kernel \cite{getreuer2011linear} remain coarse approximations of the \textit{de-facto} "fabrication kernel". ~ \figref{fig:motivation-example} summarizes the results of DOEs optimized and evaluated at combinations of different settings, suggesting that 1) hologram can only be well reconstructed when train and evaluate under the same setting; 
2) interpolation kernels play a critical role in high-frequency hologram feature construction, where a mismatched kernel can result in drastic quality decline. 
As such, this simple yet informative experiment emphasizes the importance of the "fabrication kernel" and thus motivates the need of fabrication-aware optimization. 

\begin{figure}[t]
    \centering
    \includegraphics[width=0.99\linewidth]{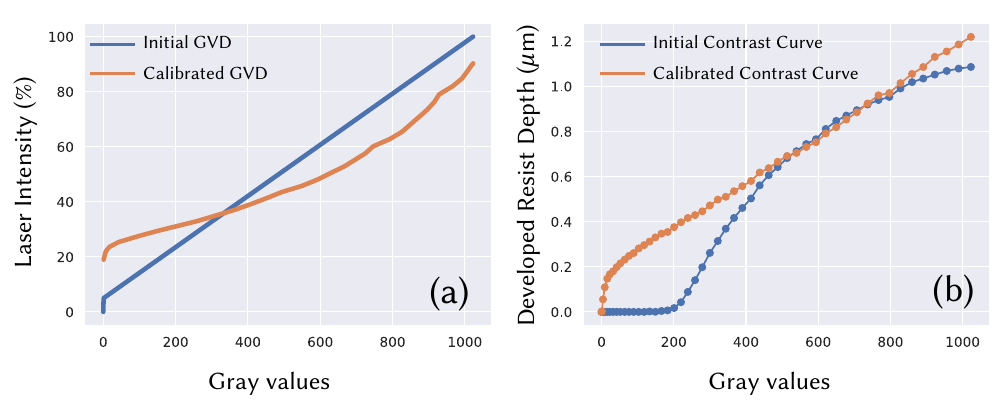}
    \caption{\textbf{Contrast Curve Calibration.} (a) the initial and calibrated gray value distribution (GVD) for intensity control of the direct-write grayscale lithography machine. A nonlinear mapping between gray values (1024 levels) and the laser intensity is utilized after calibration. (b) the initial and calibrated contrast curve of the photoresist. After calibration, we find the developed resist depth is almost perfectly linearly proportional to the gray values.   }
    \label{fig:cc-calib}
\end{figure}

\paragraph{Fabrication Pipeline.~} 
% The employed fabrication pipeline is illustrated in \figref{fig:teaser} and \ref{fig:image-formation}. 
Our fabrication pipeline utilizes the advanced direct-write grayscale lithography \cite{grushina_direct-write_2019} followed by replication with nanoimprint lithography \cite{barcelo_nanoimprint_2016}, with a step-by-step illustration in \figref{fig:teaser} (insets A-D).
% The fabrication of a diffractive optical element (DOE) using direct-write grayscale lithography followed by nanoimprint replication involves a multi-step process. 
Initially, a Soda-lime substrate is coated with a positive AZ® 4562 photoresist. A Heidelberg Instruments DWL 66+ mask writer is used in the direct-write grayscale lithography stage. A laser beam, modulated to vary intensity, selectively exposes the photoresist to create a continuous, three-dimensional relief pattern corresponding to the DOE design. The exposed photoresist is then developed, removing material proportional to the exposure dose, resulting in a smooth, multilevel surface profile. The resulting master template in the resist is then used in nanoimprint lithography: a thin layer of ultraviolet (UV)-curable resin is applied to a new substrate, and the master is pressed into it to transfer the pattern. UV light cures the resin, solidifying the replicated DOE structure. The master is then released, leaving a high-fidelity replica. This process enables scalable production of large-area DOEs. 

This fabrication pipeline offers several advantages over the conventional multi-level binary-mask-based photolithography process, which was widely used in DOE manufacturing in literature \cite{khonina_exploring_2024}.  Unlike the binary-mask approach, which requires multiple photolithography steps with separate masks for discrete height levels, grayscale lithography uses a single laser exposure to create continuous 3-D relief profiles, significantly reducing process complexity and eliminating alignment errors. 
The simplified process allows for faster prototyping and design iterations, shortening the prototyping cycle from one week to a few hours for a centimeter-scale DOE design, according to our experience in manufacturing.

% Additionally, the nanoimprint replication step allows for rapid, cost-effective mass production of DOEs with high fidelity. 
Next, we describe a two-stage calibration for the proposed super-resolved neural lithography model, a differentiable digital twin of the fabrication process.

\paragraph{Contrast Curve Calibration.~}
% To fully characterize the intrinsic "kernel" of the fabrication process, we employ a two-stage calibration method, which consists of a contrast curve calibration to rectify the nonlinear photo-response of photoresist. 
% The contrast curve indicates the relationship between the 
The input layout to the direct-write grayscale lithography system is a pixelated grayscale image with discrete 1024 levels (10 bits), whose values are monotonically mapped to relative laser intensities (0-100\%) for exposure control. This mapping, coined gray value distribution (GVD) is realized via a lookup table (LUT) that is by default a linear mapping with a small offset (See blue curve in \figref{fig:cc-calib} (a)). With pre-determined laser power and development time\footnote{Laser power and development time should be pre-determined via a trial-and-error process in order to produce desirable maximum depth, corresponding to $2 \pi$ phase modulation at nominal 550-nm wavelength, at maximum laser intensity}, the contrast curve, delineating the relationship between gray values and developed resist depths, can be obtained by creating and measuring a test pattern, which consists of a series of uniform patches, each assigning a incrementally increasing gray value.  We use a test pattern with $7 \times 7$ uniform patches, and the developed structure is then measured by a Zygo optical profilometer (NewView
7300). As shown in \figref{fig:cc-calib} (b), the default GVD leads to a nonlinear contrast curve owing to the nonlinear photo-response of the photoresist. In practice, a linear contrast curve is always preferred as it allows for the maximum dynamic range of the design space, and simplifies the downstream neural lithography learning. By numerically inverting the measured contrast curve with linear interpolation, we obtain a new calibrated GVD that yields an almost linear contrast curve (orange one in \figref{fig:cc-calib} (b)) by updating the system LUT accordingly. Some uncorrected nonlinearities remains in the low-end of the calibrated contrast curve, as a result, we exclude these gray values, and only use 24 to 1023 gray values (1000 levels in total) in design.

\begin{figure}[t]
    \centering
    \includegraphics[width=0.99\linewidth]{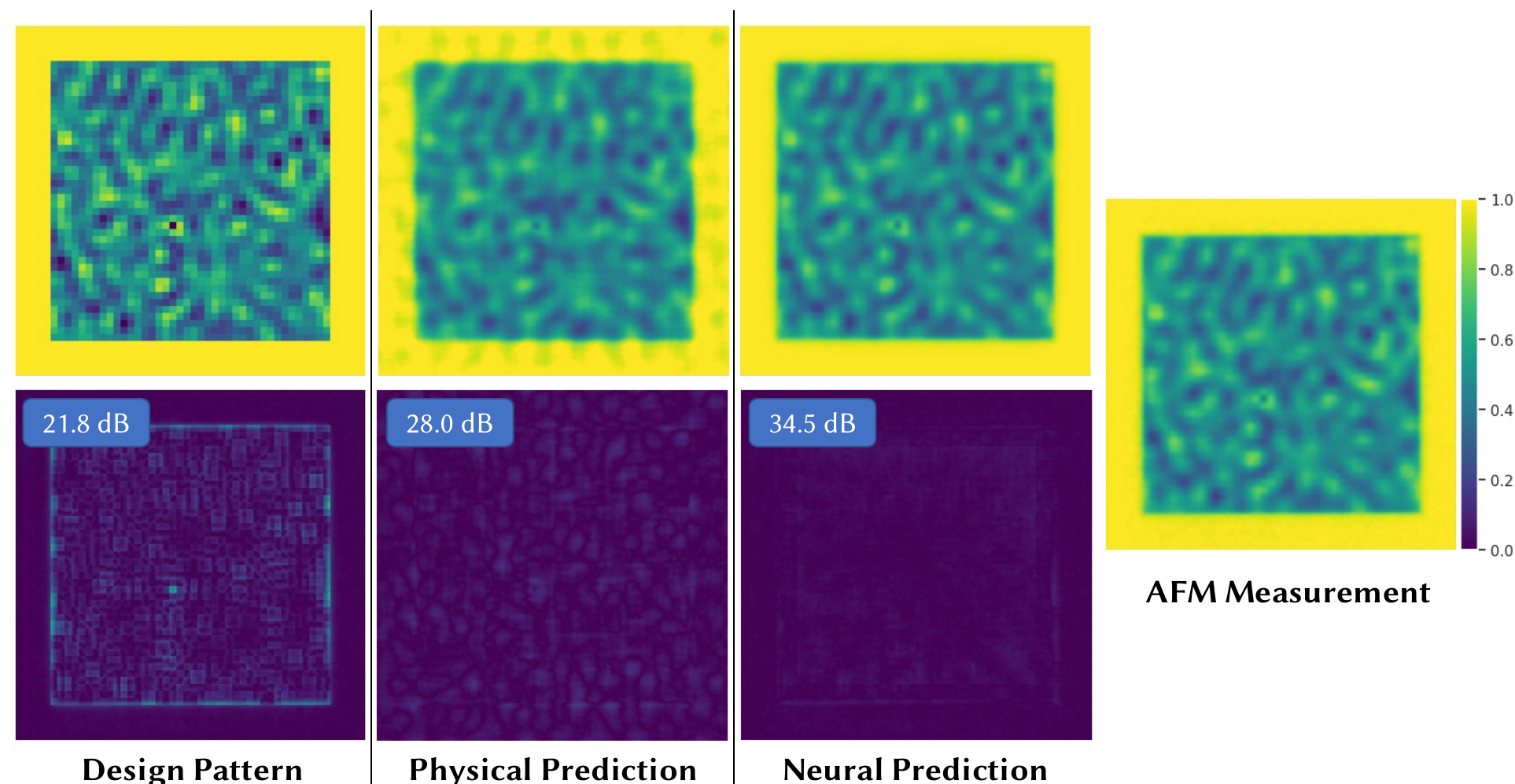}
    \caption{\textbf{Lithography Model Evaluation}.  We report here a design pattern and its corresponding AFM measurement from the constructed evalset, along with lithography model predictions---the modulation transfer function-based physical model and the proposed neural lithography model. The error maps (with respect to the AFM measurement ground truth) are also annotated with the associated PSNR values, validating the proposed model predictions.}
    \label{fig:neural-litho-calib}
\end{figure}

\begin{table}[!t]
\centering
\caption{Quantitative evaluation of two lithography models on the collected evalset. Our neural lithography model does not suffer from overfitting, while outperforming the physical model by a large margin.}
\vspace{-2mm}
\begin{tabular}{lcc}
\toprule
& \small Physical \cite{xu_fabrication-integrated_2025} & \small Neural (Ours) \\
\midrule
% Train set & 27.66 dB  & 35.27 dB \\
PSNR $\uparrow$ & 27.36 dB  & 35.20 dB  \\
NRMSE $\downarrow$ & 7.27 \%  & 2.45 \% \\
\bottomrule
\end{tabular}
\label{tab:litho-model-eval}
\end{table}

\begin{figure*}[!t]
    \centering
    \includegraphics[width=0.95\linewidth]{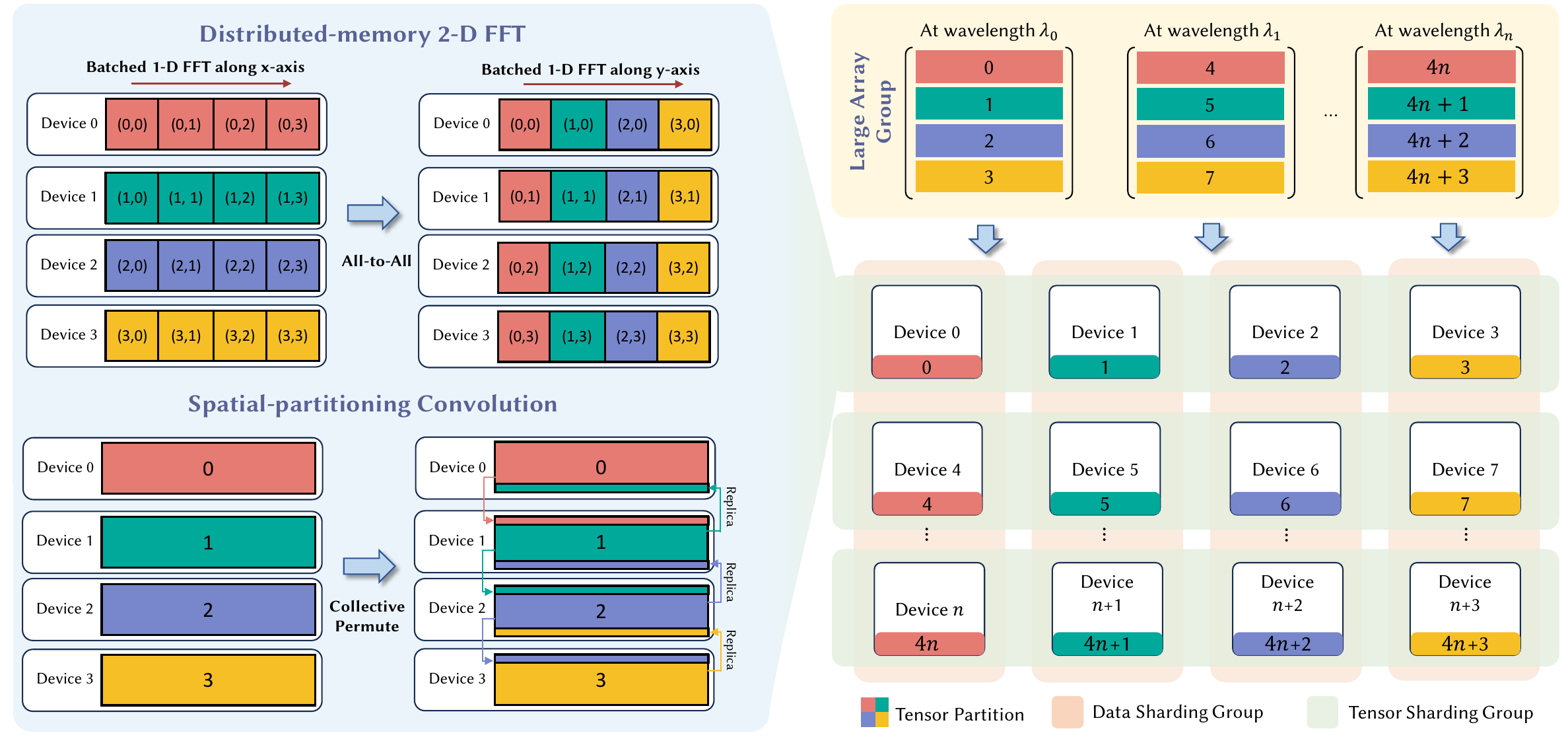}
     \caption{\textbf{GSPMD-based Distributed Computing Framework} tailored for large-area fabrication-aware diffractive optics. We illustrate the distributed computation of the proposed D$^2$FFT (top-left) and the spatial-partitioning convolution (bottom-left) leveraging tensor parallelism. The (GPU) processors can be arranged into a multi-dimensional mesh to enable arbitrary combinations of hybrid data and tensor parallelism. }
    \label{fig:tensor-parallel}
\end{figure*}

\paragraph{Neural Lithography Learning.~}

After contrast curve calibration, we design and fabricate another set of calibration patterns (spatially arranged in a single wafer) to construct the dataset for training and evaluation of the neural lithography model. Similar to prior work \cite{zhao2023close}, we randomly generate a set of 2-D patterns following a uniform distribution. A low-pass filter is then applied in the Fourier domain to limit the high-frequency components. $2\times10$ patterns in total are created with incrementally decreasing maximum cutoff frequency, each of which features \SI{40}{\micro\metre}$^2$ area discretized at \SI{1}{\micro\metre}-spacing grid (\textit{c.f.,} \figref{fig:neural-litho-calib}).
Once fabricated, the resulting sample is then measured using an atomic
force microscope (AFM) \cite{giessibl_advances_2003} to obtain precise
3-D profiles of the fabricated patterns, which can be used as ground
truth for training and evaluating the neural lithography model. For
each pattern we scan an area of
\SI{50}{\micro\metre}$\times$\SI{50}{\micro\metre}, with
256$\times$256 sampling points, resulting in approximately
\SI{200}{\nano\metre} sampling resolution. To cope with the AFM
imaging artifacts inherent in the measurement process
\cite{ricci2004recognizing}, we measure each pattern twice in two
orthogonal scanning directions. The resulting raw measurements are
then preprocessed, registered and fused to obtain final 3-D profiles
with high fidelity. 
% Detailed procedures and considerations for AFM measurements are described in the \textit{Supplemental Material}.
%In principle, more data samples are preferred to mitigate the potential risk of overfitting for deep learning models. 

We note the AFM measurement process is time-consuming and labor-intensive\footnote{Each scanning of a single pattern takes more than 20 minutes and the whole process cannot be fully automated as random errors might occur during the scanning.}---each sample might require repeated measurements due to unpredictable random errors occurring during the process, for example, the AFM probe tip might hit tiny dust atop the sample surface on occasion, leading to a failed measurement. Fortunately, we empirically find that only 10 high-quality data pairs are sufficient to learn a robust neural lithography model with good generalizability\footnote{At the early development stage of this work, we collected more AFM measurements to train the neural lithography model, but later found it was redundant and unnecessary.}.  

We split the collected 20 data pairs into training and evaluation sets, each of which has 10 data pairs. Then we construct a small yet effective convolutional neural network (CNN) to learn the mapping from design layouts (at \SI{1}{\micro\metre}-spacing grid) to their corresponding AFM measurements (\SI{200}{\nano\metre}-spacing grid), which amounts to 5$\times$ super resolution. As shown in \figref{fig:image-formation}, the CNN consists of a few convolutional layers (2 to 4 layers) with ReLU nonlinearity followed by a pixel shuffle layer \cite{shi_real-time_2016} for upsampling. Skip connections are also employed to prevent the gradient vanishing problem \cite{he_deep_2016}, which is especially important since the gradient flow must be appropriately propagated to the model input (design layout) for fabrication-aware optimization. 
% We train a sequence of models with different complexities for distinct upsampling ratios (\ie, 1$\times$, 2$\times$, 4$\times$, 5$\times$) where the ground truth is obtained by resampling high-resolution AFM measurements with linear interpolation. 
% More implementation details can be found in the \textit{Suppl. Material}.

We train the CNN as our fabrication surrogate using the Adam optimizer with an initial learning rate of $10^{-3}$ that decays to zero in 3000 iterations following a cosine schedule. Stochastic optimization with mini-batch size of 5 and random data augmentation such as horizontal/vertical flipping and rotations are adopted for regularizing model learning to avoid overfitting. We also calibrate a modulation transfer function (MTF) based physical model \cite{xu_fabrication-integrated_2025} using the collected trainset, by fitting a MTF that transfers nearest upsampled design patterns to their AFM measurements in Fourier domain. Qualitative and quantitative evaluations of the fitted lithography models are provided in \figref{fig:neural-litho-calib} and Table \ref{tab:litho-model-eval} respectively, suggesting the effectiveness of the proposed neural lithography model for capturing structure-dependent "fabrication interpolation kernel" which is otherwise difficult to be modeled by the MTF-based linear physical model. We note that more complex NN architectures like the one used in \cite{zhao2023close} (with an extra upsampling head) can be employed, which may achieve comparable or better results compared to the proposed simple CNN. However, we find that further reducing testing errors is meaningless considering the intrinsic random variations in both fabrication and measurement processes.% Besides, more complex architectures are prone to overfitting given a small amount of training data, and incur higher computational costs in downstream fabrication-aware optimization.

\section{Parallel Computing for Large-Scale Wave Simulation}
\label{sec:parallelization}

The design framework detailed in the previous section introduces very
large simulation grids that can accurately represent small-scale DOE
features at wavelength scale. Unfortunately, while accurate, this pipeline also drastically
increases the memory requirements for forward simulation and
especially for inverse design, rendering large-scale design tasks
infeasible on existing single-GPU pipelines due to the limited amount
of GPU memory available.

We address this bottleneck by developing parallel computing tools that
can distribute for example large FFT computations across GPUs (potentially located at multiple nodes).

\subsection{Large-scale Wave Propagation via D$^2$FFT}
\label{sec:dist-fft}

Per the ASM in \eqref{eq:asm}, the FFT lies at the heart of the (Fourier) wave optics, which turns out to be an inevitable bottleneck for high-resolution large-scale wave optics simulation---the loaded memory exceeds the GPU memory capacity which makes the model unable to fit entirely on a single GPU. To address this issue, we implement a D$^2$FFT, following the spirit of tensor parallelism. 

Multidimensional FFT can be efficiently computed by
a sequence of lower-dimensional FFTs. For instance, a 2-D FFT can be performed by first applying a 1-D FFT along the rows, followed by a 1-D FFT along the columns. Such a row-column decomposition naturally leads to a distributed-memory FFT implementation with an additional communication step. 
% As shown in \figref{fig:tensor-parallel}, a large 2-D array is partitioned into smaller sub-arrays, each of which is assigned to a different GPU. 
To execute a large-scale 2-D FFT, we first partition the given 2-D array into multiple sub-array chunks along the columns (Y-axis), each of which is assigned and allocated to a different (GPU) device. Under this arrangement, the row (X-axis) information is completely preserved in each device, and hence, a batched 1-D FFT along rows (X-axis) can be performed independently on each device.
Then, a MPI-style primitive for collective communication, \textit{all-to-all} communication \revise{\cite{snir1998mpi,doi2010overlapping,alabed_partir_2025}}, is invoked to exchange unique chunks of the distributed array between participating devices. Mathematically, this is analogous to a matrix transpose operation, such that the resulting array is now partitioned along the rows (X-axis), which is followed by a batched 1-D FFT along the Y-axis on each device to complete the 2-D FFT. Note the spatial sharding axis of the distributed array swaps after performing this distributed-memory 2-D FFT. The whole process is illustrated in \figref{fig:tensor-parallel} (top-left). We implement this operation in Jax \cite{bradbury2018jax} framework in a differentiable way, where the backward pass is realized by another distributed-memory 2-D FFT since the discrete Fourier transform can be represented by a symmetric matrix. Differentiable distributed-memory inverse FFT can be implemented similarly. 

\subsection{Tensor-parallel Convolutional Neural Network}
\label{sec:tpcnn}
With D$^2$FFT and thereby large-scale ASM in hand, we are almost ready to address the optimization problem in \eqref{eq:fab-aware-opt} at scale. However, the computational load of the neural lithography model $\mathcal{G}$ also becomes intractable for large-scale DOE designs. 
Despite the simplicity of the CNN we adopted as the neural lithography model, it still requires a large amount of GPU memory in dealing with high-resolution large-area design layouts (\eg, a \SI{1}{\centi\metre}$^2$ input layout at \SI{1}{\micro\metre}-spacing grid is discretized as a 10,000$\times$10,000 array as the CNN's input) which is a rather unusual case in DL. %It worth noting that 
The common sliding-window strategy to circumvent this issue is only applicable at inference, since the fabrication-aware optimization requires the objective gradient to be backpropagated to the input layout at training. As such, akin to the D$^2$FFT, tensor parallelism must be leveraged to make the CNN scalable to large-sized inputs. 

To perform the tensor-parallel spatial-partitioning convolution, we split the input array into multiple sub-arrays along the Y axis\footnote{The spatial sharding axis should be chosen to be compatible with the following D$^2$FFT to avoid unnecessary data resharding.}. Since the convolution (in DL) is a local operation, each sub-array can be independently processed by the same convolutional kernel on each device, except for processing the sub-array boundary. To ensure consistent results around the boundary, each sub-array must be padded with the corresponding boundary information from neighboring devices, thus requiring communications. This special communication operation can be efficiently implemented by another MPI-style primitive, coined \textit{collective permute}. By executing this operation, each sub-array has $(\frac{H}{N} + 2\times\floor*{\frac{K}{2}}) \times W$ elements with replicated boundary values, where $H$, $W$ are the height and width of the global intact array, $N$ is the number of devices, and $K$ denotes the kernel size (or receptive field in general) of the convolutional layer, $\floor*{\cdot}$ indicates the floor function.
An illustration of this spatial-partitioning convolution is shown in \figref{fig:tensor-parallel} (bottom-left).

% Given a sub array of size $\frac{H}{N} \times W$, where $N$ is the number of devices, the padding size $P$ (in one side) is determined by $P = \floor*{\frac{K}{2}}$ where $K$ is the kernel size (or receptive field in general) of the convolutional layer.

\begin{figure}[!htbp]
\begin{minted}[
  frame=lines,
  framesep=2mm,
  baselinestretch=1.2,
  fontsize=\footnotesize
]{python}
"""
A multi-dimensional processor mesh is created with 2x2 devices 
where first and second axes are named wavelength ("wvl") 
and tensor parallel ("tp") respectively.
"""
mesh = MeshShardingHelper([2, 2], ['wvl', 'tp'])
@partial(
    mesh.sjit, 
    args_sharding_constraint=(
        PartitionSpec('wvl', 'tp', None), # for "field"
        PartitionSpec('wvl', None, 'tp'), # for "H"
    ),
    out_shardings=PartitionSpec('wvl', 'tp', None), # for "E_prop"
)
def propagate(field, H):
    U = fft_func(field)  # Perform 2D FFT (implemented elsewhere)
    E_k_prop = U * H
    E_prop = ifft_func(E_k_prop)  # Perform 2D inverse FFT
    return E_prop
\end{minted}
\caption{\textbf{Example GSPMD Segment} for the free-space wave propagator. An 2$\times$2 processor mesh is created to realize hybrid data and tensor parallelism. Pure data or tensor parallelism can be achieved by simply setting the mesh size to [4, 1] or [1, 4], respectively. All tensors have three dimensions (axes), and the sharding annotations/constraints suggest the spatial sharding axis (for tensor parallelism) of "field" and "H" are the Y and X axes, respectively, because the D$^2$FFT would swap this sharding axis from Y to X.}
\label{fig:example-gspmd}
\end{figure}

% \begin{figure}[!htbp]
% \begin{lstlisting}
% """
% A multi-dimensional processor mesh is created with 2x2 devices 
% where first and second axes are named wavelength ("wvl") 
% and tensor parallel ("tp") respectively.
% """
% mesh = MeshShardingHelper([2, 2], ['wvl', 'tp'])
% @partial(
%     mesh.sjit, 
%     args_sharding_constraint=(
%         PartitionSpec('wvl', 'tp', None), # for "field"
%         PartitionSpec('wvl', None, 'tp'), # for "H"
%     ),
%     out_shardings=PartitionSpec('wvl', 'tp', None), # for "E_prop"
% )
% def propagate(field, H):
%     U = fft_func(field)  # Perform 2D FFT (implemented elsewhere)
%     E_k_prop = U * H
%     E_prop = ifft_func(E_k_prop)  # Perform 2D inverse FFT
%     return E_prop
% \end{lstlisting}
% \caption{\textbf{Example GSPMD Segment} for the free-space wave propagator. An 2$\times$2 processor mesh is created to realize hybrid data and tensor parallelism. Pure data or tensor parallelism can be achieved by simply setting the mesh size to [4, 1] or [1, 4], respectively. All tensors have three dimensions (axes), and the sharding annotations/constraints suggest the spatial sharding axis (for tensor parallelism) of "field" and "H" are the Y and X axes, respectively, because the D$^2$FFT would swap this sharding axis from Y to X.}
% \label{fig:example-gspmd}
% \end{figure}

\subsection{GSPMD Implementation}
\label{sec:implementation}

Handcrafting computational diffractive optics systems with hardware-associated operators from above is often time-consuming and error-prone. To enable flexible configurations (such as arbitrary combinations of data and tensor parallelism for any hardware topology) we implement the framework in Jax following the GSPMD programming model, leveraging an automatic, compiler-based parallelization system \cite{xu2021gspmd,alabed_partir_2025}. The essential abstractions of GSPMD are a multi-dimensional mesh of processors and a sharding annotation system. The former arranges the available computing devices into a multi-dimensional mesh with named sharding axes, while the latter annotes the partition specifications (PartitionSpec) of the input and output tensors of a given function. A simplified example of a GSPMD implementation segment of the ASM is shown in \figref{fig:example-gspmd}. By utilizing this programming model and the XLA compiler \cite{openxla2025}, the hardware-specific MPI-style primitive is automatically generated and invoked in the compiled program free of error-prone hard encoded low-level MPI calls. 
% Examples of compiled code segments involving generated low-level MPI calls can be found in Supplementary Material.

\begin{figure}[!t]
    \centering
    \includegraphics[width=0.99\linewidth]{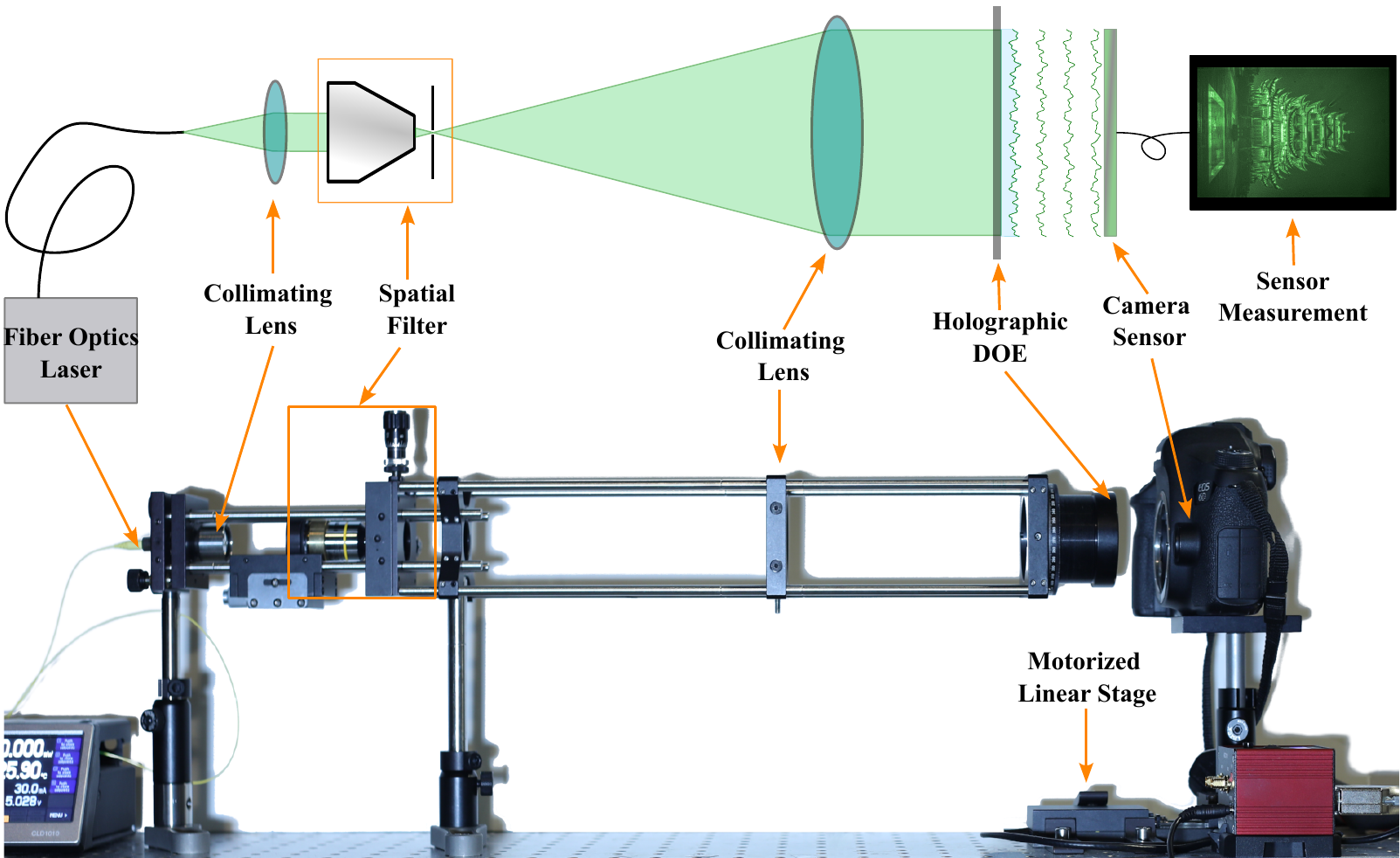}
    \caption{\textbf{Experimental Holographic Setup.} We validate the fabrication-aware DOE designs with the experimental setup shown above. Here, a collimated laser beam is modulated by the DOE, which then propagates to the photosensor directly to form the designed hologram image. }
    \label{fig:holo-setup}
\end{figure}

\begin{figure*}[!t]
    \centering
    \includegraphics[width=0.99\linewidth]{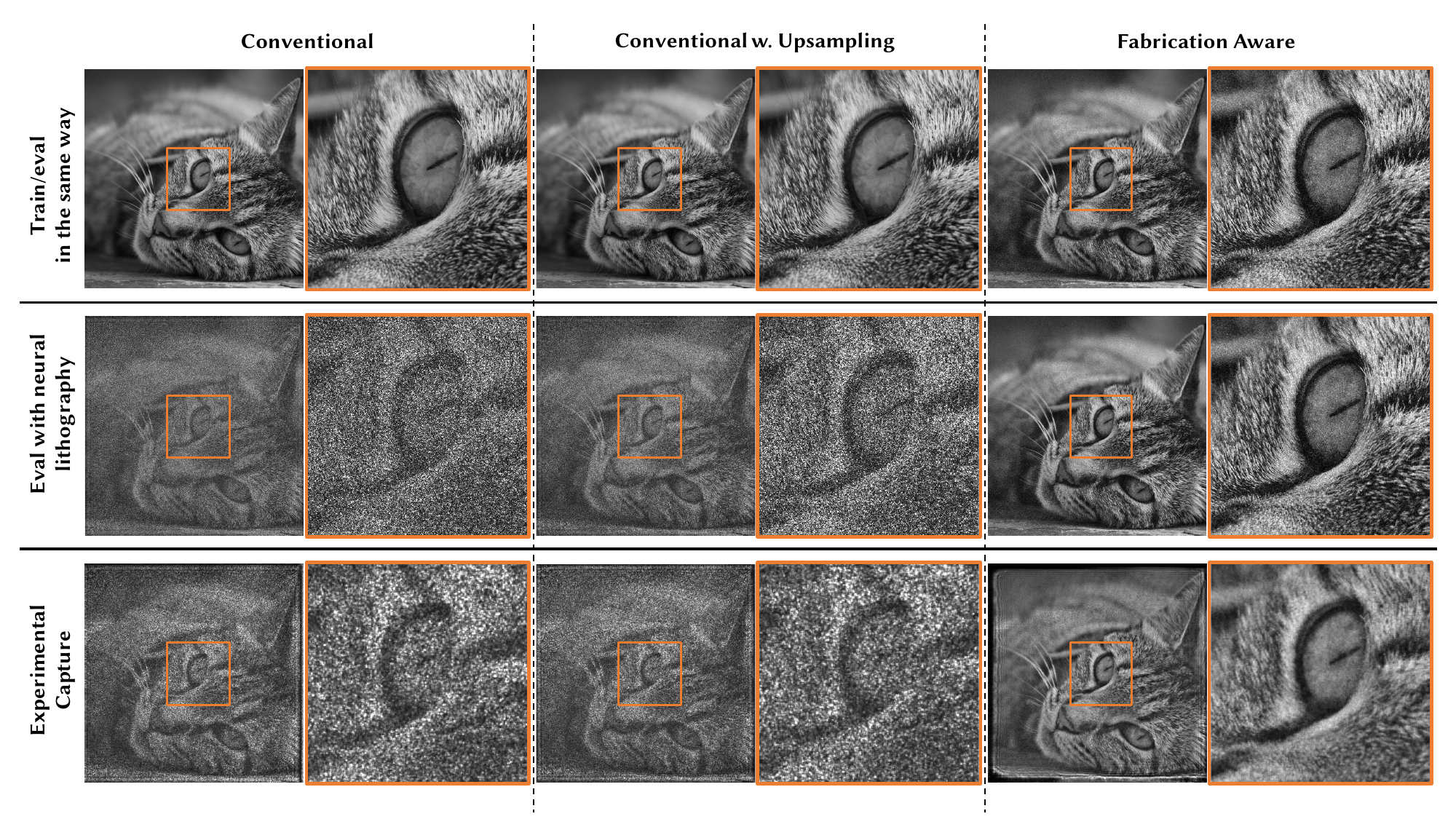}
    \caption{\textbf{Experimental and Synthetic Evaluation of Proposed Fabrication-aware CHDs}. In simulation, 1) all approaches are successful when trained and evaluated under the same setting (row 1); 2) the quality of the conventional approaches declines drastically under the neural lithography setting (row 2) which matches the experimental captures (row 3). The fabricated DOE designed through our approach generates almost speckle-free, high-resolution coherent hologram close to the simulation, without any additional optical filtering.}
    \label{fig:holo2d-cat}
\end{figure*}

\section{Applications and Analysis}

In this section, we evaluate the proposed fabrication-aware design method for several applications, including computational display holography (Section \ref{sec:display-holography}), beam shaping (Section \ref{sec:beam-shaping}), and single-DOE broadband color imaging (Section \ref{sec:broadband-imaging}). For all applications, we validate the effectiveness of our approach not only in simulation but also with an experimental prototype, resulting in 11 fabricated DOEs in total (6 for holography display, 2 for beam shaping, 3 for broadband imaging). The readers are encouraged to review the video in the Supplementary Material.

\begin{figure*}[!t]
    \centering
    \includegraphics[width=0.99\linewidth]{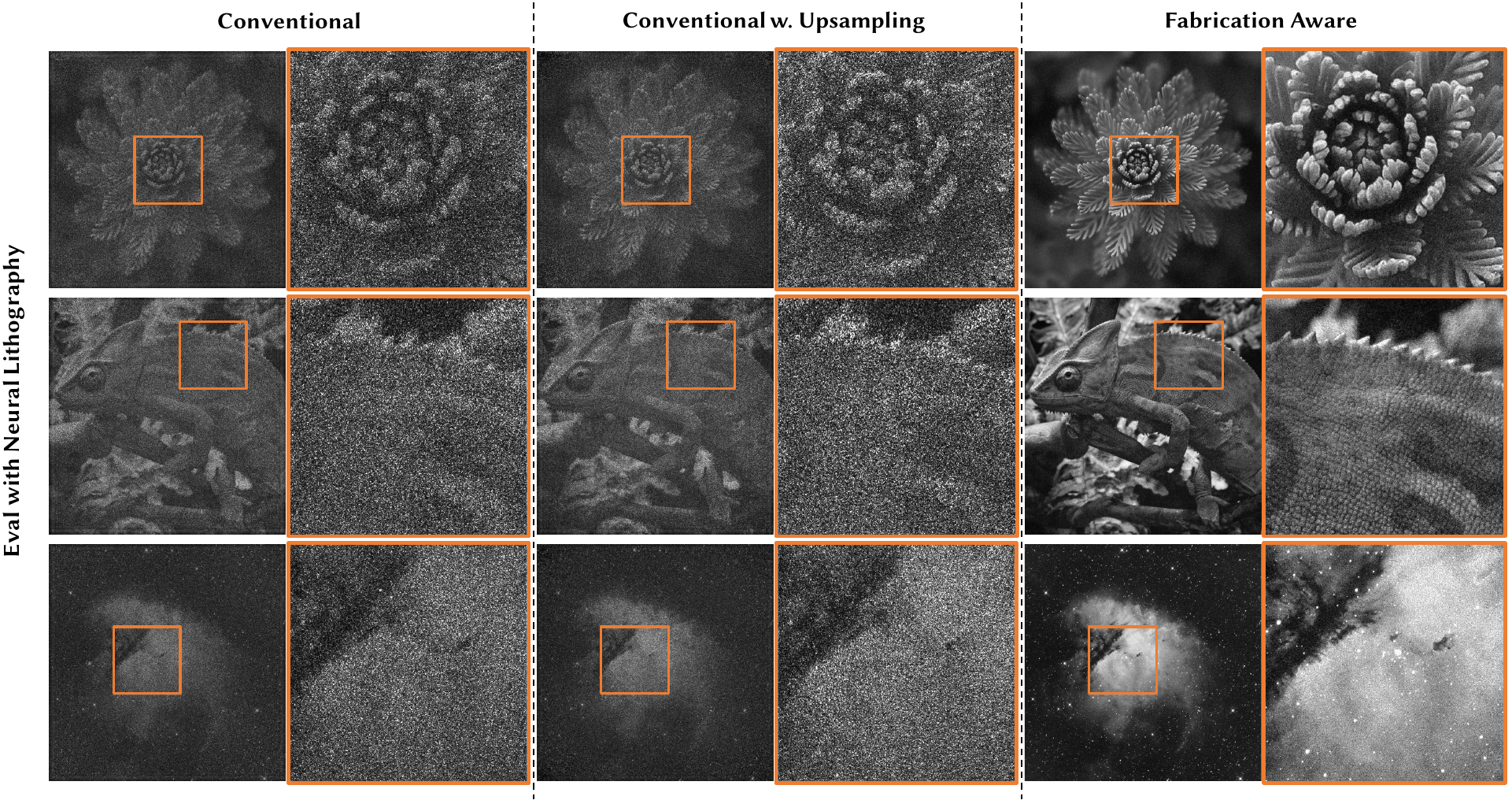}
    \caption{\textbf{Simulation CHD Results for Diverse Scenes.} For all of the scenes, our fabrication-aware approach consistently produces high-resolution, clean holograms with complex details that are barely visible in conventional approaches.}
    \label{fig:holo2d-others-sim}
\end{figure*}

\subsection{Prototype Fabrication}

% \todo{Brief paragraph of fabrication process and mounting for applications here.}

We follow the fabrication pipeline from Sec.~\ref{sec:neural-lithography} and fabricate microstructured patterned DOEs using direct-write grayscale lithography with the Heidelberg Instruments DWL 66+ mask writer, employing AZ® 4562 photoresist on soda-lime glass to create smooth 3D relief structures in a single exposure step. The process involves spin-coating, soft-baking at 120°C, writing with 1023 grayscale levels, and developing with AZ® 726 MIF for 25 seconds. These patterns are then transferred into OrmoComp, a UV-curable polymer, via room-temperature nanoimprint lithography using the Obducat Eitre 3 system, where the photoresist mold is pressed into the polymer, cured with UV light, and released to produce high-fidelity micro-optical components. After fabrication, the DOE wafers are diced and mounted in the respective experimental setup described in the corresponding sections below. 
% \kaixuan{Since we have two set of different experimental setup, I would describe them seperately in their own subsections.}

\subsection{Computational Holographic Display}
\label{sec:display-holography}

Computational holographic display (CHD) \cite{chakravarthula_wirtinger_2019,peng_neural_2020} employs algorithms to simulate optical hologram creation and reconstruction. The computed holographic image is brought to life through a display, usually featuring an illuminating source and a phase-modifying element (DOEs). We focus on 2-D near-field computational holography. Specifically, we use Adam optimizer to solve the optimization problems \eqref{eq:general-opt-obj} and \eqref{eq:fab-aware-opt}, where the loss function is defined in \eqref{eq:hologram-loss}, together with the training recipe in Section \ref{sec:neural-lithography} (Toy Example). Three \revise{3.6$\times$3.6 mm$^2$} DOEs with \SI{2}{\micro\metre} feature size are designed and fabricated to generate the same 2-D \revise{3.6$\times$3.6 mm$^2$} hologram image in 1 cm away from the DOE, for which we build the experimental setup as shown in \figref{fig:holo-setup}. The optics in our setup are designed to produce a smooth, spatially clean, and evenly collimated beam (520.6-nm wavelength). This is achieved by using a spatial filter as an intermediate step between expander stage. The resulting beam enables plane-wave illumination for holograms of various sizes. Our physical realization does not require any additional intermediate image plane (such as those implemented with a 4-$f$ system to filter out undesirable diffraction orders \cite{gopakumar_unfiltered_2021}), thus directly assessing the patterns diffracted from DOEs. Finally, for accurate sensor positioning, we employed a motorized linear stage with a positioning precision of \SI{10}{\micro\metre}.
We evaluate \revise{3.6$\times$3.6 mm$^2$} DOEs with three different design approaches\footnote{\revise{Extra results of Xu et al. \citeyearpar{xu_fabrication-integrated_2025} and our fabrication-aware approach with 2$\times$ neural upsampling are provided in the Supplementary Material.}}:
\begin{itemize}
    \item \textit{Conventional}: conventional differentiable optics optimization solving \eqref{eq:general-opt-obj} at \SI{2}{\micro\metre}-spacing grid without upsampling; 
    \item \textit{Conventional w. Upsampling}: conventional design from above \eqref{eq:general-opt-obj} at \SI{250}{\nano\metre}-spacing grid with $8\times$ nearest upsampling;
    \item \textit{Fabrication-aware}: proposed fabrication-aware approach of \eqref{eq:fab-aware-opt} at \SI{250}{\nano\metre}-spacing grid with $8\times$ neural upsampling. 
\end{itemize}
All these DOEs 
can be efficiently optimized under a single A-100 GPU (\textit{c.f.,} the compute specifics in Table \ref{tab:compute-specifics}), however, to realize fabrication-aware optimization of larger-area DOE, tensor-parallel computing routines must be utilized. 
\figref{fig:holo2d-cat} summarizes both the simulation and experimental results for these holograms. Remarkably, our fabrication-aware approach results in an experimental realization of a nearly speckle-free, high-definition cat hologram under the coherent laser illumination. As such, we find that it largely closes the design-to-manufacturing gap evidenced by conventional approaches. We note the residual minor contrast and resolution difference between simulation and experimental results are likely attributed to the imperfection of the experimental setup itself, such as the inexact 3-D printed aperture (as evidenced by the diffraction patterns around boundary), the glass, and anti-aliasing filters atop the photosensor. Additional computer-generated hologram results (in simulation) can be found in \figref{fig:holo2d-others-sim}, which confirm the effectiveness of the proposed method for realizing diverse and vivid hologram images with abundant details.

\begin{table*}[!t]
\centering
\caption{\textbf{Compute Cost for Optimizing Task DOEs.} The discrete grid characterizes the discretization of the transfer function of ASM (as well as the super-resolved DOE). The 2-D GSPMD mesh size indicates the number of devices assigned to data and tensor parallel dimension. For CHD and broadband imaging, we optimize the DOEs for 3,000 and 20,000 iterations, respectively, for which we report the total compute time of the optimization process. 
}
\vspace{-2mm}
\begin{tabular}{lccccc}
\toprule
 DOE &  Design Approach &  Discrete Grid &  \# of GPUs & GSPMD Mesh &  Compute Time  \\
\midrule
3.6mm $\times$ 3.6mm  & Conventional & 3,600$\times$3,600  & 1  & [1, 1] & 42 s \\
CHD \& Beam Shaping & Conventional w. Upsampling & 28,800$\times$28,800  & 1  & [1, 1] & 14 m \\
(\textit{c.f.,} \figref{fig:holo2d-cat}, \ref{fig:beam-shaping})  & Fabrication Aware & 28,800$\times$28,800  & 1  & [1, 1] & 17 m \\
\midrule
32.16mm $\times$ 21.44mm & Conventional w. Upsampling & 128,640$\times$85,760  & 16  & [1, 16] & 13 h  \\
CHD (\textit{c.f.,} \figref{fig:holo2d-temple}) & Fabrication Aware & 128,640$\times$85,760  & 16  & [1, 16] & 15 h \\
\midrule
10mm $\times$ 10mm & Conventional w. Upsampling & 6$\times$40,000$\times$40,000  & 12  & [6, 2]  & 9 h \\
Imaging (\textit{c.f.,} \figref{fig:broadband-imaging}) & Fabrication Aware & 6$\times$40,000$\times$40,000  & 12  & [6, 2]  & 10 h \\
\bottomrule
\end{tabular}
\label{tab:compute-specifics}
\end{table*}

\begin{figure*}[!t]
    \centering
    \includegraphics[width=0.99\linewidth]{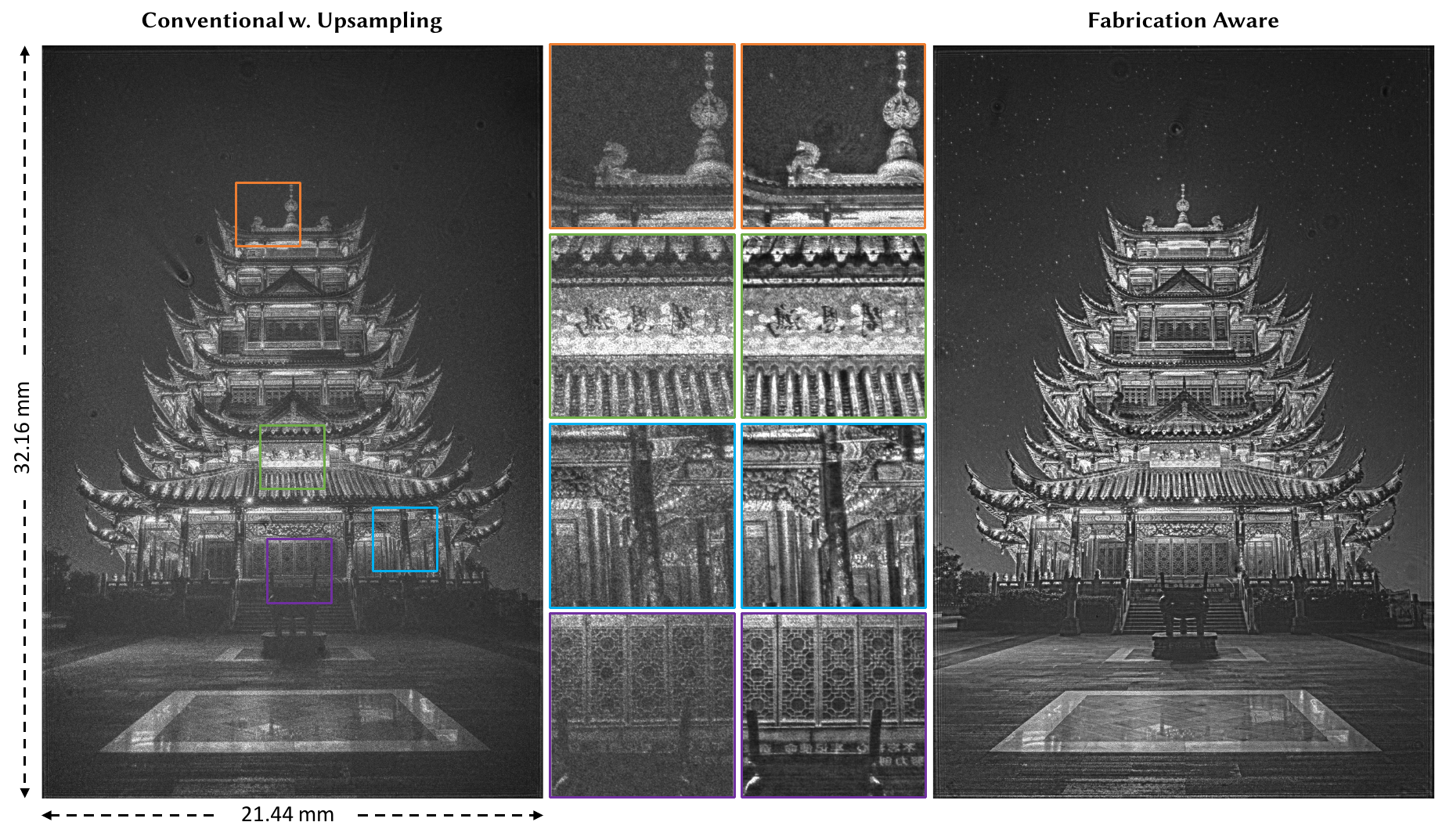}
    \caption{\textbf{Experimental, Large-area, High-definition (4K $\times$ 6K) 2-D Hologram Reconstruction}.  Two 32.16 mm $\times$ 21.44 mm DOEs with \SI{2}{\micro\metre} feature size are designed and optimized at \SI{0.5}{\micro\metre}-spacing grid (4$\times$ super resolution) using conventional (with nearest upsampling) and our fabrication-aware approaches, distributed across 16 A-100 GPUs through tensor parallel. We compare the resulting holograms side-by-side in real-world experiments, where our fabrication-aware approach generates sharper, clearer, high-SNR details than the conventional one. }
    \label{fig:holo2d-temple}
\end{figure*}

\begin{figure}[!t]
    \centering
    \includegraphics[width=0.99\linewidth]{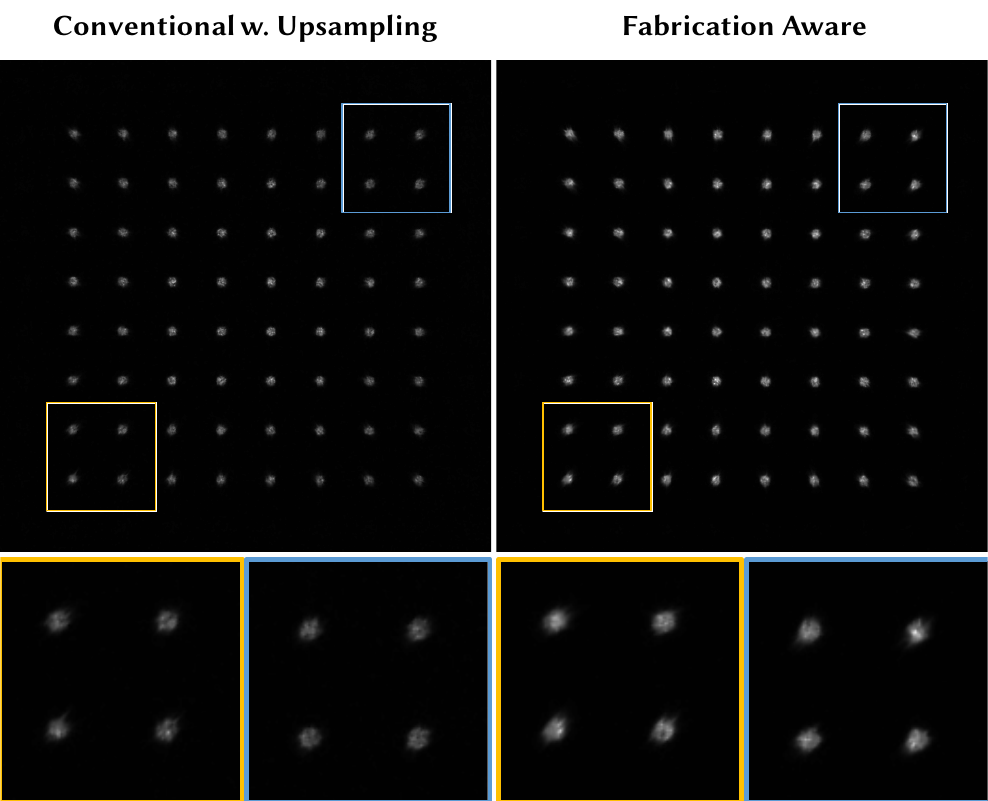}
    \caption{\textbf{Experimental Evaluation of Diffractive Beam Splitters.} We show the experimental raw measurements captured under the same lighting condition, where we keep the exposure time, laser power constant for both measurements. The intensity of measurements directly reflects the relative diffraction efficiency of the beam splitters, validating the effectiveness of the fabrication-aware beam splitter device. }
    \label{fig:beam-shaping}
\end{figure}

\begin{figure}[!t]
    \centering
    \includegraphics[width=0.99\linewidth]{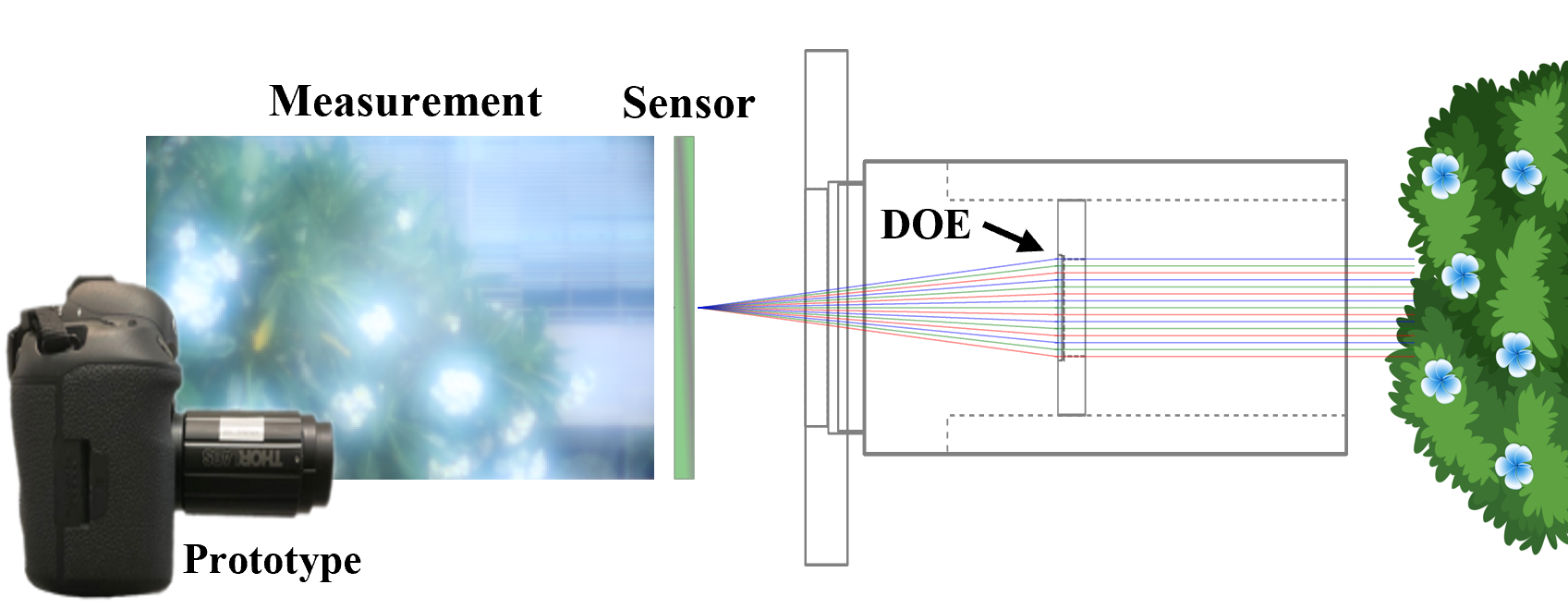}
    \caption{\textbf{Single-DOE Broadband Imaging Setup.} For the proposed imaging application, install the designed DOE inside a telescopic lens tubes to assemble the f/8 imaging setup with design focal length. We mount the tube with an adapter on an off-she-shelf DSLR camera (left).}
    \label{fig:imaging-setup}
\end{figure}

\paragraph{Large-Area DOEs for Ultra-definition Near-field Holography}
To confirm the scalability of the proposed tensor-parallel framework tailored for large-area DOE designs, we further optimize two \SI{2}{\micro\metre} feature-sized DOEs with 32.16 mm in height and 21.44 mm in width (corresponding to 4K $\times$ 6K pixels in the CanonEOS5D5 sensor we used), discretized at a \SI{0.5}{\micro\metre}-spacing grid (4$\times$ super-resolution). These DOEs are optimized to produce the same-sized holograms at 6 cm away from the DOEs, given the plane-wave coherent illumination. 
This optimization task results in large-scale intermediary arrays with 128,640 $\times$ 85,760 pixels beyond the memory capacity of a single A-100 GPU (even a single computing node of 8 A-100 GPUs). We instead employ our GSPMD-based distributed computing framework, and deploy the DOE optimization into two nodes consisting of 16 A-100 GPUs, which readily fit this huge array spatially shard onto 16 segments. The experimental measurements (using the holographic setup in \figref{fig:holo-setup}) for both designs (conventional and ours) are shown in \figref{fig:holo2d-temple}, which clearly validates the scalability of our proposed approach in handling large-area DOE designs at an unprecedented scale.

\subsection{Beam Shaping with Diffractive Beam Splitter}
\label{sec:beam-shaping}

Beam shaping, as one of the classic use cases of DOE \cite{kress2000digital} has found downstream applications in laser material processing \cite{kuang2013ultrafast}, fast LiDAR 3-D sensing \cite{yuan2021fast} and visual vibrometry \cite{zhang2023impacts}. 
Here, we design DOEs as diffractive beam splitters whose goal is to split the collimated laser beam into a regular grid of beams that yields an array of spots with flat-top intensity profiles at the desired plane. This task can be viewed as a special case of computer-generated hologram---instead of generating an image, we steer the DOE to produce an array of focal spots in the destination plane. We find that beam splitter designs without considering the fabrication deformation inevitably lose diffraction efficiency and the spot array intensity uniformity. 
Again, we design two \revise{3.6$\times$3.6 mm$^2$} DOEs with \SI{2}{\micro\metre} feature size, steering the incident plane-wave beam into a 8$\times$8 array of flat-top equal-intensity spots in a focal plane located at 1 cm away from the DOE, by minimizing the loss function in \eqref{eq:hologram-loss}.
We reuse the experimental setup of the holographic display (\figref{fig:holo-setup}) and measure the raw intensity patterns for fabricated beam-splitting DOEs under the same lighting condition for fair comparison of diffraction efficiency. 
% We also capture a background image without installing the DOE at the same exposure setting, which is used to characterize the absolute diffraction efficiency of these diffractive beam splitter. 
The experimental measurements of diffracted spot-array patterns are exhibited in \figref{fig:beam-shaping}, where the overall spot intensity of the fabrication-aware design is \textbf{53\%} higher than the conventional one, indicating large diffraction efficiency gains brought by our method.  

\begin{figure*}[!t]
    \centering
    \includegraphics[width=0.99\linewidth]{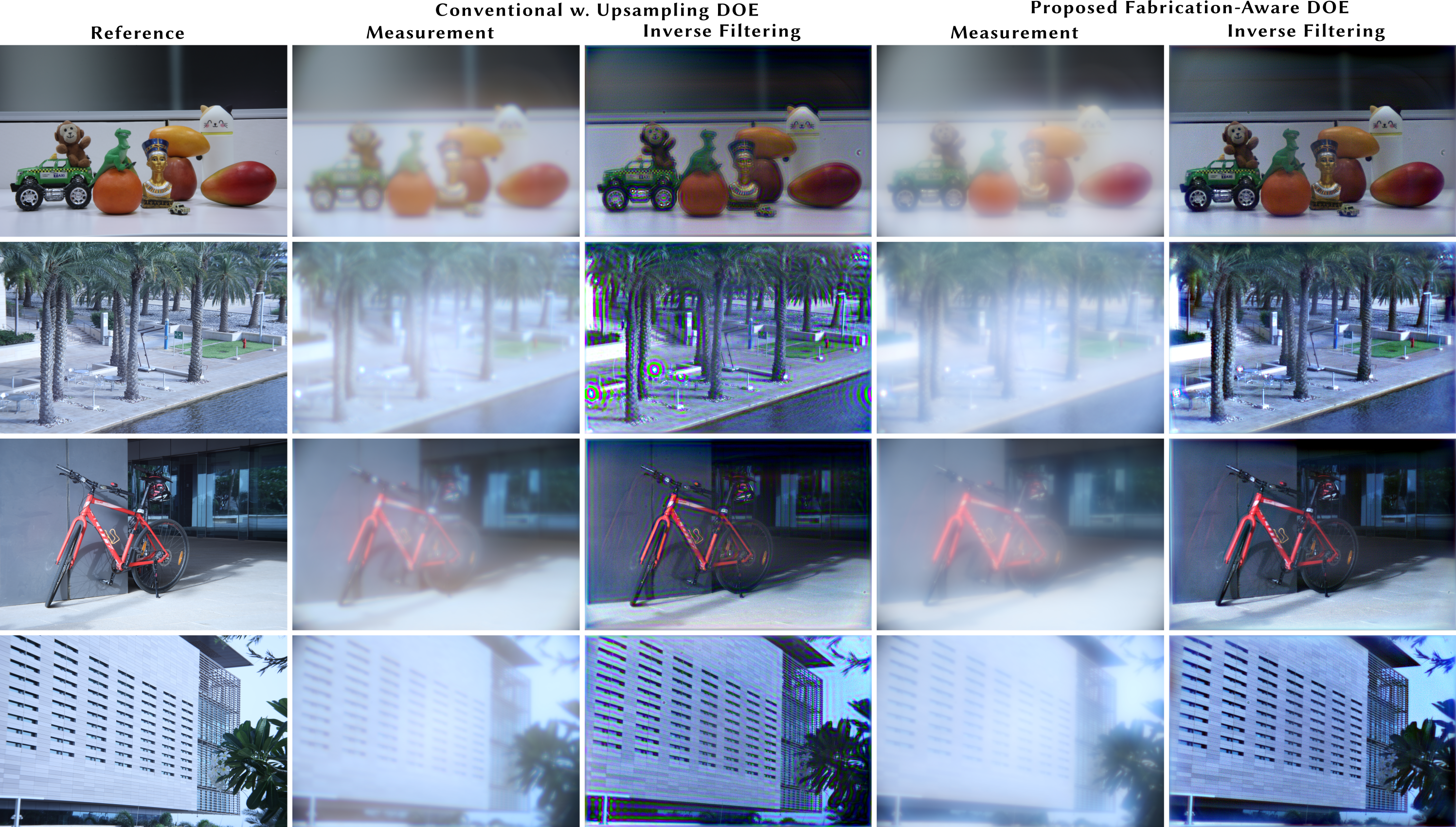}
    \caption{\textbf{Experimental Validation of Fabrication-aware Broadband DOE for Imaging.} We capture diverse indoor and outdoor scenes with both conventional nearest up-sampling DOE and proposed fabrication-aware DOE (column 1, 2, and 4). To visualize the discrepancy between the designed PSF and fabricated PSF, we conduct a Wiener filtering using simulated PSF on the experimental captured scenes assuming high SNR ($\gamma = 5 \times {10}^{-4}$). Our proposed fabrication-aware design yields high-fidelity results, while the conventional nearest up-sampling results in images with severe chromatic aberration, ringing artifacts, and elevated noise. }
    \label{fig:broadband-imaging}
\end{figure*}

\subsection{Single-DOE Broadband Color Imaging}
\label{sec:broadband-imaging}

Broadband color imaging with diffractive optics systems is fundamentally challenging due to strong wavelength-dependent dispersion \cite{peng_diffractive_2016, aieta2015multiwavelength}. High-fidelity broadband reconstruction demands point-spread functions (PSFs) that are simultaneously invertible and spectrally consistent across densely sampled bands \cite{peng_diffractive_2016, sun2025collaborative, froch2025beating}, which notably increases the memory requirements of the design optimization. In addition, discrepancies between simulated and fabricated PSFs further degrade image quality \cite{shi2024learned, chakravarthula_thin_2023}. We validate that the distributed fabrication-aware method addresses both issues.

Using the GSPMD implementation described in Section~\ref{sec:implementation}, we shard the DOE model across two devices, sample 6 wavelengths uniformly over the visible range (400 nm to 700 nm) leveraging hybrid data and tensor parallelism (with GSPMD mesh size $[6, 2]$). We enforce radial symmetry on the phase profile by optimizing a 16th-order polynomial radial vector. Following the evaluation of the previous applications, two DOE variants (conventional with nearest upsampling and our fabrication-aware approach with neural upsampling) are trained with the same broadband imaging loss

\begin{equation}
\label{eq:broadband_loss}
\mathcal{L}_{\mathrm{imaging}} = \mathcal{L}_{\mathrm{focus}}
+ \mathcal{L}_{\mathrm{consistency}}
+ \mathcal{L}_{\mathrm{energy}}
\end{equation}

Here, $\mathcal{L}_{\mathrm{focus}}$ drives the PSF to concentrate its energy at the designated center pixel, ensuring a sharp focal peak; $\mathcal{L}_{\mathrm{consistency}}$ penalizes variations in PSF shape across the sampled wavelengths, enforcing spectral uniformity; and $\mathcal{L}_{\mathrm{energy}}$ maximizes the total PSF energy to avoid trivial, zero‐energy solutions. 
For more details, please refer to the Supplementary Materials (Section B.3).

We fabricated both DOE variants and integrated them into the imaging rig as shown in \figref{fig:imaging-setup}. A diverse set of indoor and outdoor scenes were captured under various illumination conditions to quantitatively and qualitatively assess the performance of each lens design. A representative subset of scenes can be found in Fig.~\ref{fig:broadband-imaging}, column 1.

To illustrate the simulation–reality discrepancy, we first convert the captured raw image into linear-RGB image space with details in Supplementary Materials (Section B.4). We then apply Wiener deconvolution \eqref{eq:wiener_filter} to each \emph{captured} linear–RGB measurement $y_{\mathrm{cap}}$ using its corresponding \emph{simulated} on-axis PSF $k_{\mathrm{sim}}$, that is

\begin{equation}
\label{eq:wiener_filter}
\hat{x}
=
\mathcal{F}^{-1}\!\Biggl\{
\frac{\overline{K_{\mathrm{sim}}}}{\bigl|K_{\mathrm{sim}}\bigr|^2 + \gamma}
\;Y_{\mathrm{cap}}
\Biggr\}.
\end{equation}

Here, capitalization denotes the Frequency domain version of the variables, and \(\overline{(\cdot)}\) denotes complex conjugation.  

When the fabricated PSF closely matches the design as for proposed fabrication-aware DOE (Fig.~\ref{fig:broadband-imaging}, column 4 and 5), the deconvolved reconstruction $\hat{x}$ closely approximates the ground truth. By contrast, the conventional, nearest-upsampling DOE (Fig.~\ref{fig:broadband-imaging}, columns 2 and 3) exhibits PSF mismatches that manifest as wavelength-dependent chromatic aberrations, elevated noise in poorly invertible spectral bands, and ringing artifacts under the high-SNR inversion assumption ($\gamma = 5\times10^{-4}$). These results confirm the effectiveness of our fabrication-aware optimization in closing the simulation-to-reality gap.

We also remark that the image quality of our neural designs
is already competitive with existing single-DOE computational imaging
systems, despite the basic single-step reconstruction method \revise{with simulated PSFs} employed
here. We find that lightweight, edge-friendly reconstruction methods can be used when an accurate optical design process is taken into account, enabled by our fabrication-aware pipeline. 
Additional experimental results, \revise{including the comparisons of the inverse-filtering results using simulated and experimental PSFs}, can be found in Supplementary Materials (Section C.2).

\section{Discussions}

This section discusses the proposed framework in terms of scalability, effectiveness, and limitations.

\begin{figure}[!t]
    \centering
    \includegraphics[width=0.99\linewidth]{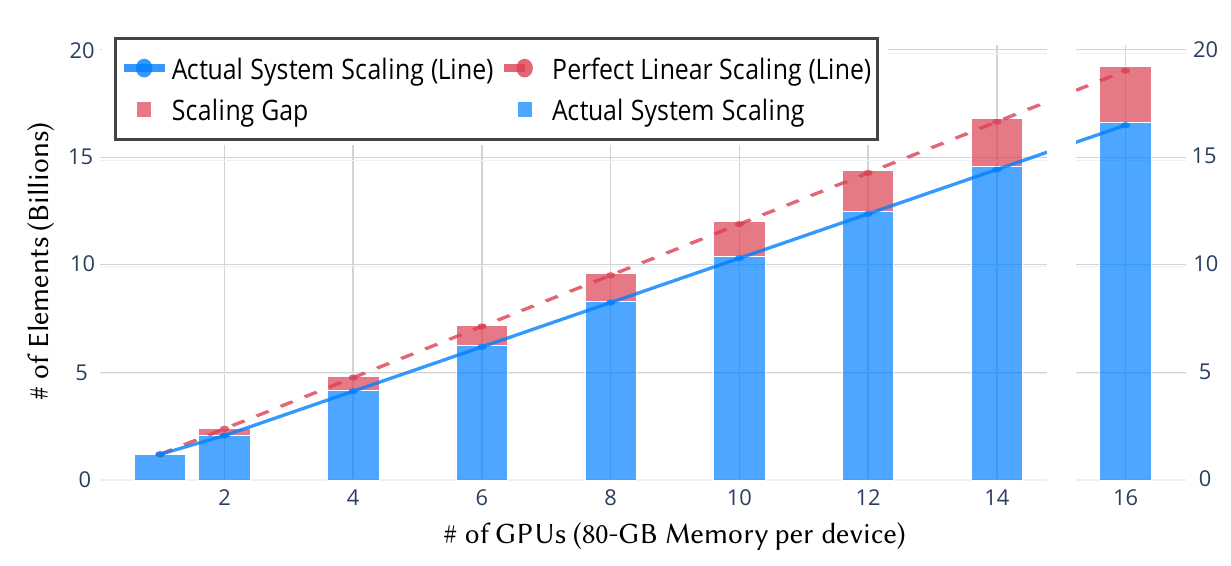}
    \caption{\textbf{Scaling Analysis} of the tensor-parallel computing routines (D$^2$FFT and spatial-partitioning convolution) using the coherent computer-generated hologram as the evaluation benchmark. We report here the largest system (indicated by the number of elements of the discrete grid) scaling with the number of devices (80-GB memory per device).   }
    \label{fig:scaling}
\end{figure}

\paragraph{Scaling Analysis of the Tensor-parallel Algorithms} 
% Though we have demonstrated the optimization of large-area a few centimeter-scale DOEs at sub-micron resolution leveraging tensor parallelism, 
We conduct a scaling analysis of the proposed tensor-parallel computing routines for large-area computational diffractive optics. 
We take the coherent computer-generated hologram as the benchmark, and test the largest system size (the total number of elements of the discrete grid of a super-resolved DOE) that can be fitted in a given number of GPU devices. The analysis is reported in \figref{fig:scaling}, where we evaluate multiple device configurations from 1 to 16. We find sub-linear scaling (1.73$\times$ rather than perfect 2$\times$ linear scaling) from 1 to 2 devices, which is expected because of the communication overhead. Our method scales linearly from 2 to 16 GPU processors (up to the available computing resources we access to), validating scalability of our distributed computing framework. 

%This is obvious and a design choice.
%One aspect of our tensor-parallel system is the efficiency evaluation, however, we note the D$^2$FFT is not a compute-efficient algorithm since the \textit{all-to-all} communication cost overshadows the multi-processing speedup. Deliberately distributing a FFT computation into multiple GPU devices often results in lengthier processing time.  As a result, the tensor-parallel D$^2$FFT should always be used for combating memory bottleneck rather than speeding up.  

% \paragraph{Conventional/Neural OPC vs. Fabrication-aware Optimization}

\begin{figure}[!t]
    \centering
    \includegraphics[width=0.99\linewidth]{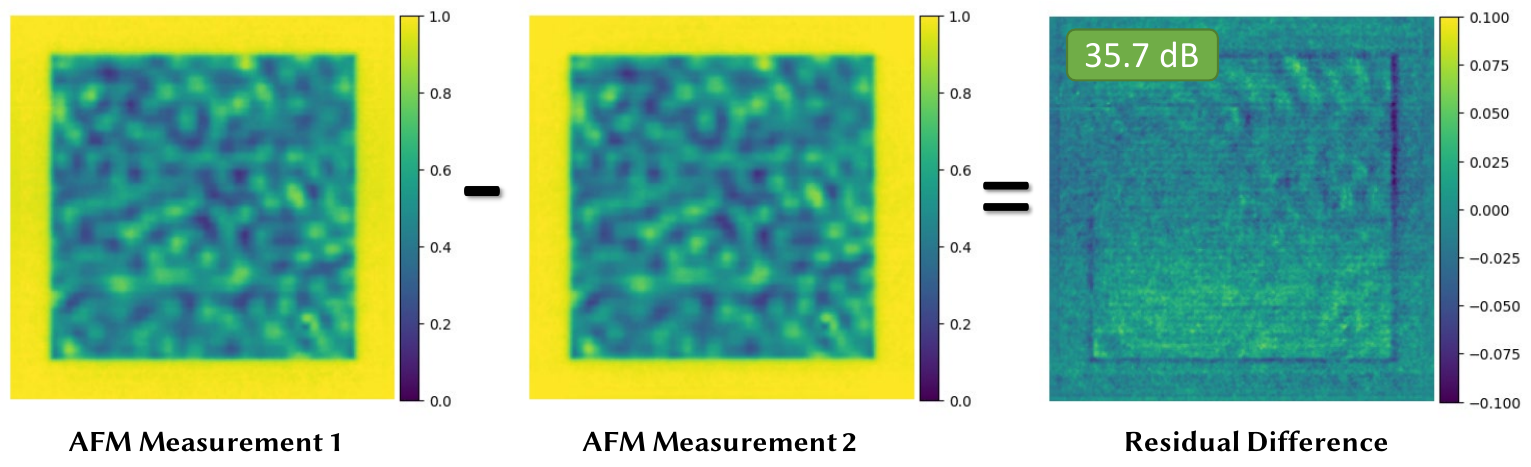}
    \caption{\textbf{Fabrication Uncertainty Analysis} of the neural lithography pipeline. Two identical design patterns are fabricated repeatedly. Analyzing these two patterns post-fabrication yields slightly different AFM measurements, which inherently limits the precision upper bound of the neural lithography model. }
    \label{fig:data-uncertainty}
\end{figure}

\paragraph{Limitations on Model/Data Uncertainty}
We note that the randomness of the fabrication and measurement processes inherently limits our neural lithography model's precision (forward predictability). To analyze this limitation for our specific process, in preparation for the calibration patterns used to train the neural lithography model (Section \ref{sec:neural-lithography}), we deliberately place multiple identical patterns in the design layout periphery. Ideally, these patterns post-fabrication would be identical; however, as shown in \figref{fig:data-uncertainty} for two patterns, slight differences due to random fabrication and measurement error remain, setting the upper bound of prediction accuracy of the neural lithography model.
\revise{Nevertheless, a significant portion of the observed random variation stems from our current academic fabrication facility, which relies on manual interventions such as the control of development time---introducing operator-dependent variability. In commercial foundries, such processes are fully automated with substantially reduced operational variance and higher repeatability.}

\revise{
\paragraph{Model Generalizability}
The proprietary and diverse nature of nano/micro-fabrication recipes makes it infeasible to directly apply a model trained in one foundry to another, or even within the same foundry using different materials. Nevertheless, our methodology is universal and can be adapted to other lithography processes through appropriate calibration, rendering it highly suitable for clean-room training. Currently, AFM measurements represent the primary bottleneck in the calibration workflow; however, this requirement remains manageable, as the process can be completed within one day using only 10 relatively small patches.}

\revise{
\paragraph{Camera-in-the-loop (CITL) Approaches}
Unlike CITL approaches that learn a \textit{global} mapping for wave propagation, our digital twin of the manufacturing is designed to learn \textit{local} shape deviations between the design and the fabricated device.
These local variations correspond to process-induced effects such as optical blur, nonlinear material response, and 3D chemical reactions. 
Our model does not account for global effects. Based on our extensive experience with DOE fabrication, we find such global variations—for instance, those related to position on the wafer—are minimal. This sets nano-fabrication apart from digital holography works employing spatial light modulators (SLM) that may suffer voltage gradients across the chip, and other non-uniformities.
The focus on locality offers practical advantages: it allows the model to be trained with limited data and enables rapid recalibration for new material systems or process conditions. In contrast, a full CITL pipeline (for static DOE) with potentially hundreds of prototypes would be intractable in practice, as are AFM measurements for large area samples required to train global models.
}

\section{Conclusion}
This work tackles two limitations that existing computational diffractive optics struggle with: 1) the design-to-manufacturing gap and 2) the inability to simulate and optimize large-area devices. To this end, we propose a fabrication-aware, end-to-end optimization method with a super-resolved neural lithography model as a differentiable fabrication surrogate. To design large area optics, we develop a tensor-parallel algorithm (\ie, D$^2$FFT and spatial-partitioning convolution), which overcomes problem-inherent memory limitations and supports large-scale wave propagation and diffractive optics optimization at sub-micron resolution. 
We validate our method with direct-write grayscale lithography and nanoimprint replication, a process employed today for mass production of large-area devices. Experiments across computational display holography, beam shaping, and broadband color imaging validate that the method can produce high-quality diffractive optical systems effectively.

We believe our contributions offer a step toward bridging the gap between simulation and mass-market implementation of differentiable imaging optics. Moving forward, incorporating additional fabrication variables, such as material-specific variations, direct-write laser beam, 
%or exploring fabrication-in-the-loop strategies, 
would be interesting areas for research. We believe these advancements provide a foundation for broader adoption of learned optical systems in real-world imaging and display applications.

\begin{acks}
 We thank Dongyu Du and Arturo Burguete Lopez for their supports.  
 Wolfgang Heidrich acknowledges support from KAUST Individual Baseline Funding.
 Felix Heide was supported by an NSF CAREER Award (2047359), a Packard Foundation Fellowship, a Sloan Research Fellowship, a Disney Research Award, a Sony Young Faculty Award, a Project X Innovation Award, a Bosch Research Award and an Amazon Science Research Award. Fabrication was carried out in the Nanofabrication CoreLabs (NCL) at KAUST. 
 
\paragraph{Author contributions:} K.W., W.H. and F.H. conceived the ideas; K.W. designed, implemented and analyzed the system and framework; H.J. and K.W. built the physical setup and captured the experimental results; H.A. established the fabrication pipeline and produced all designed devices; J.S. performed the image reconstruction for the imaging task; K.W., W.H. and F.H. led the manuscript
writing; H.J., H.A., J.S., Q.F. assisted in design and analysis, experiments, and manuscript writing; W.H. and F.H. supervised the project.
\end{acks}

\bibliographystyle{ACM-Reference-Format}
\bibliography{bibliography}

\end{document}